\RequirePackage[2020-02-02]{latexrelease}
\documentclass[nofootinbib,aps,letterpaper,preprintnumbers,superscriptaddress,showkeys]{revtex4}
\usepackage{amsmath}
\usepackage{eurosym}
\usepackage{amssymb}
\usepackage{graphicx}
\usepackage{color}
\usepackage{braket}
\usepackage{graphicx,epsfig}
\usepackage{amsmath}
\usepackage{amssymb}
\usepackage{subfigure}
\usepackage{amsfonts}
\usepackage[left=1.5cm,right=1.5cm,top=1.5cm,bottom=1.5cm]{geometry}
\usepackage[usenames,dvipsnames,svgnames]{xcolor}  
\usepackage{hyperref}   
\definecolor{beamer@PRD}{RGB}{46,48,146}
\hypersetup{
	breaklinks=true,
	pdfstartview={FitH},    
	colorlinks=true,       
	linkcolor=blue,          
	citecolor=purple,        
	filecolor=cyan,      
	urlcolor=cyan,           
	anchorcolor=pink,      
	linktocpage=true
}

\begin{document}
	\date{\today}
	\newcommand{\ve}{\varepsilon}
	\newcommand{\be}{\begin{eqnarray}}
		\newcommand{\ee}{\end{eqnarray}}
	\newcommand{\bea}{\begin{eqnarray}}
		\newcommand{\eea}{\end{eqnarray}}
	\newcommand{\bbra}[1]{\mbox{$\left\langle\, #1 \right\mid$}}
	\newcommand{\bket}[1]{\mbox{$\left\mid #1\,\right\rangle$}}
	\newcommand{\pro}[2]{\mbox{$\langle\, #1 \mid #2\,\rangle$}}
	\newcommand{\expec}[1]{\mbox{$\langle\, #1\,\rangle$}}
	\newcommand{\expecl}[1]{\mbox{$\left\langle\,\strut\displaystyle{#1}\,\right\rangle$}}
	\newcommand{\real}{\mbox{{\rm I\hspace{-2truemm} R}}}
	\renewcommand{\natural}{\mbox{{\rm I\hspace{-2truemm} N}}}
	\renewcommand{\a}{\hat a}
	\renewcommand{\H}{\hat H}
	\renewcommand{\b}{\hat b}
	\renewcommand{\v}{\hat v}
	\renewcommand{\d}{\mbox{{\rm d}}}
	\def\comment#1{}
	\def\mn#1{*\marginpar[{*{\scriptsize #1}}]{*{\scriptsize #1}}}
	\def\Re{{\rm Re\,}}
	\def\Im{{\rm Im\,}}
	\def\mbf#1{#1}
	\newcommand{\re}{\real{\rm e}}
	\renewcommand{\a}{\hat a}
	\newcommand{\ac}{\hat a^{\dagger}}
	\renewcommand{\b}{\hat b}
	\newcommand{\bc}{\hat b^\dagger}
	\newcommand{\lp}{\ell_{\rm p}}
	\newcommand{\mpl}{m_{\rm p}}
	\newcommand{\gn}{G_{\rm N}}
	\newcommand{\rh}{r_{\rm H}}
	\newcommand{\Rh}{R_{\rm H}}
	\renewcommand{\d}{\mbox{${\rm d}$}}
	\newcommand{\ep}{\mathcal{E}_{\rm p}}

	\title{Probing the Lorentz Symmetry Violation Using the First Image of Sagittarius A*: Constraints on Standard-Model Extension Coefficients }

	\author {\textbf{Mohsen Khodadi}}
	\email{m.khodadi@ipm.ir}
	\affiliation{School of Astronomy, Institute for Research in Fundamental Sciences (IPM),	P. O. Box 19395-5531, Tehran, Iran}
	\affiliation{Physics Department, College of Sciences, Shiraz University, Shiraz 71454, Iran}
	\affiliation{Biruni Observatory, College of Sciences, Shiraz University, Shiraz 71454, Iran}	
	
	\author{\textbf{Gaetano Lambiase}}\email{lambiase@sa.infn.it}
	\affiliation{Dipartimento di Fisica ``E.R Caianiello", Universit$\grave{a}$ degli Studi di Salerno, Via Giovanni Paolo II, 132-84084 Fisciano (SA), Italy}
	\affiliation{Istituto Nazionale di Fisica Nucleare - Gruppo Collegato di Salerno - Sezione di Napoli, Via Giovanni Paolo II, 132 - 84084 Fisciano (SA), Italy}
	
	\begin{abstract}
		Thanks to unparalleled near-horizon images of the shadows of Messier 87* (M87*) and Sagittarius A* (Sgr A*) delivered by the Event Horizon Telescope (EHT), two amazing windows opened up to us for the strong-field test of the gravity theories as well as fundamental physics. Information recently published from EHT about the Sgr A*'s shadow lets us have a novel possibility of exploration of Lorentz symmetry violation (LSV) within the Standard-Model Extension (SME) framework. Despite the agreement between the shadow image of Sgr A* and the prediction of the general theory of relativity, there is still a slight difference which is expected to be fixed by taking some fundamental corrections into account. We bring up the idea that the recent inferred shadow image of Sgr A*  is explicable by a minimal SME-inspired Schwarzschild metric containing the Lorentz violating (LV) terms obtained from the post-Newtonian approximation.
		The LV terms embedded in Schwarzschild metric are dimensionless spatial coefficients ${\bar s}^{jk}$ associated with the field responsible for LSV in the gravitational sector of the minimal SME theory. In this way, one can control Lorentz invariance violation in the allowed sensitivity level of the first shadow image of Sgr A*. Actually,  using the bounds released within $1\sigma$ uncertainty for the shadow size of Sgr A* and whose fractional deviation from standard Schwarzschild, we set upper limits for the two different combinations of spatial diagonal coefficients and the time-time coefficient of the SME, as well. The best upper bound is at the $10^{-2}$ level, which should be interpreted differently from those constraints previously extracted from well-known frameworks since unlike standard SME studies it is not obtained from a Sun-centered celestial frame but comes from probing the black hole horizon scale.
	\end{abstract}

	\maketitle

	\section{Introduction}
	
	Black holes address the most extreme regions of space-time \cite{Luminet:1998wr}, and represent one of the most important and  natural prediction of the general theory of relativity (GTR). It is widely believed that our understanding of the nature of space-time, as well as, some aspects of fundamental physics, particularly the behavior of gravity in the strong-field regime and the quantum gravity issue (the reconciliation of quantum mechanics and gravity in a unified framework), is deeply dependent on uncovering the mystery of the black hole \cite{Barack:2018yly}.  In other words, the main clue to fundamental physics is expected to come from the black hole. In the light of extraordinary advances of the technology, we have now been ushered into a golden era where black holes and their observational consequences are detectable on a wide range of scales. 
	One of impressive examples in this sense are the first images of space-time around supermassive black holes delivered by the team of Event Horizon Telescope\footnote{In this respect, it is essential to mention some leading literature related to prior observations of black holes in the galactic center by LIGO/Virgo detectors \cite{LIGOScientific:2016aoc,LIGOScientific:2016sjg,LIGOScientific:2017zid,LIGOScientific:2017bnn,LIGOScientific:2017ycc,LIGOScientific:2017vox} and the Keck Telescope/Gravity Collaboration \cite{Ghez:1998ph,Schodel:2002py,Ghez:2003qj,Eisenhauer:2005cv,Ghez:2008ms,Meyer:2012hn,GRAVITY:2020qsl,GRAVITY:2018ofz}.} (EHT).
	This exciting adventure to uncover the mystery of the deep of spacetime, in essence, began in 2019 with recording the first image of the M87* compact object \cite{EventHorizonTelescope:2019dse,EventHorizonTelescope:2019ths,EventHorizonTelescope:2019pgp,EventHorizonTelescope:2019ggy} and exposing the structure of the surrounding magnetic field \cite{EventHorizonTelescope:2021bee,EventHorizonTelescope:2021srq}. 
	In this direction, EHT's team has recently surprised us again by recording the first images of a supermassive compact object located at the center of the Milky Way galaxy i.e., Sagittarius A* (Sgr A*) \cite{EventHorizonTelescope:2022xnr,EventHorizonTelescope:2022wok,EventHorizonTelescope:2022urf,EventHorizonTelescope:2022xqj,EventHorizonTelescope:2022gsd}. 
	
	The main trait observed in these VLBI-based images, indicating the existence of a compact object such as the black hole in the universe, is a bright emission ring surrounding a central dark depression so that the latter is expected to be the black hole shadow \cite{Falcke:1999pj}. The essential condition so that the radius of the bright ring can act as a proxy for the black hole shadow radius is that the geometrically and optically emission regions are, respectively, thick and thin at the wavelength related to the VLBI network, see for example Refs. \cite{Narayan:2019imo,Bronzwaer:2021lzo}. This is exactly the case for both compact objects M87* and Sgr A*. However, there are some concerns among researchers about whether the VLBI near-horizon images related to M87* and Sgr A* address the black hole shadow surrounded by the photon ring
	\cite{Gralla:2019xty,Gralla:2019drh, Gralla:2020srx,Gralla:2020pra,Glampedakis:2021oie}. The horizonless compact objects known as black hole mimickers \footnote{There are some interesting proposals that try to distinguish between black holes and mimickers through the footprints left in gravitational waves, see \cite{Cardoso:2019nis,Maggio:2021uge,Fang:2021iyf,Cardoso:2022fbq}.} are candidates of interest which may disclose the nature of the mentioned shadow images \cite{Bambi:2008jg,Ohgami:2015nra,Shaikh:2018lcc,Shaikh:2018kfv,Joshi:2020tlq,Guo:2020tgv,Herdeiro:2021lwl} (see also review paper \cite{Cardoso:2019rvt}). Specifically, by making this assumption that maybe the object at the center of the Milky Way galaxy is a horizonless supermassive object with a surface re-emitting incident radiation, in Ref. \cite{EventHorizonTelescope:2022xqj} it has been demonstrated that the present information of Sgr A* provides strong upper bounds on the radius of such a thermal surface. However, with a closer look at the key assumption at the heart of this analysis, its results were challenged in \cite{Carballo-Rubio:2022imz}.
	It is interesting to note that the next generation of EHT, owing to a high dynamic range, will provide the possibility to penetrate the shadow of compact object  (filling the photon ring), giving a direct observation of the event horizon as the key characteristic of a black hole \cite{Chael:2021rjo}.	
	The relation between the bright ring and black hole shadow angular diameters gives us this unprecedented possibility of using black hole shadow to test fundamental physics. This is possible, especially once the black hole mass-to-distance ratio is known \cite{Johannsen:2015hib,Psaltis:2018xkc}. 
	Since the release of M87*'s shadow by EHT, many papers have addressed the corrections originating from alternative theories of gravity and new physics (see e.g. Refs \cite{Bambi:2019tjh}-\cite{Patel:2022acr})\footnote{We need to point out that due to a large number of papers on this matter and the lack of possibility to mention all of them here, these cited references are merely examples of a much more extensive collection.}. Despite extensive previous studies in the light of M87*'s shadow, rearranging them this time for the Sgr A* is well-justified and recommended in the literature
	\cite{Johannsen:2011mt,Johannsen:2015mdd,Goddi:2016qax}. In Ref. \cite{Vagnozzi:2022moj} one can find a list of reasons for evaluating the theories via Sgr A*'s shadow, in addition to its counterpart M87*.
	As a result, upon the recent release of updated and new information on the near-horizon image of Sgr A*, we are faced a spectrum of observational evaluations of different metrics \cite{Vagnozzi:2022moj}-\cite{Pantig:2022ely}.
	
	In light of the above discussion, the scope of this manuscript is to use the first image of Sgr A* to constrain the Lorentz symmetry violation (LSV). Lorentz invariance, as is well known, represents one of the fundamental symmetries in nature which both GTR and the standard model of particle physics rely on.
	Such a symmetry asserts that the outcome of any local experiment does not depend on the velocity as well as the orientation of the laboratory in which experiments are performed. However, this idea that Lorentz symmetry may not be a fundamental symmetry of nature and breaks down at some levels is supported by some fundamental theories, such as string theories \cite{Kostelecky:1989jw,Kostelecky:1988zi}, loop quantum gravity \cite{Thiemann:2002nj}, multiverses \cite{Bjorken:2002sr}, brane-world scenarios \cite{Burgess:2001vr,Frey:2003jq,Cline:2003xy}, Einstein-aether gravity \cite{Jacobson:2000xp} (see also \cite{Tasson:2014dfa,Mattingly:2005re} for other theories).
	In the last years, several tests in different fields of physics have been proposed to search for a possible breaking of Lorentz symmetry  \cite{Kostelecky:2008bfz}. Generally, one can consider three theoretical  motivations for doing these tests. First, fundamental theories based on unifying the quantum and gravitation principles  claim that the search for new physics (particularly Planck-scale physics) is strongly dependent on disclosing the mystery of Lorentz symmetry. Second, due to the deep relation between Lorentz symmetry and CPT (charge, parity, and time-reversal), the symmetry allows to predict the behavior of antimatter by investigating the behavior of matter. Third, in non-gravitational experiments, Lorentz symmetry is part of the Einstein Equivalence Principle (EEP). 
	If LSV occurs, the equations of motion may change depending on whether the experiment is boosted or rotated relative to a background field. The EEP, as the foundation of GTR and all metric theories of gravity, implies that the spacetime metric minimally couples to all forms of matter \cite{Tino:2020nla}, meaning that  EEP breaks down if the matter part of the action is modified. It is worth noting that the shadow image of M87 *  has made it possible to perform some tests on the EEP \cite{Yan:2019hxx,Li:2021mzq}.
	
	From a theoretical point of view, having an effective field theory is essential for describing observable signals of LSV. In this regard, Kostelesky and collaborators have developed a well-established effective field theory, the so-called Standard Model Extension (SME), which incorporates all possible violations of the Lorentz invariance \cite{Colladay:1996iz,Colladay:1998fq}. More precisely, it extends the standard model of particle physics and GTR by including all possible Lorentz-violating (LV) terms that can be constructed at the level of the Lagrangian. In fact, the characterization of low-energy effects of Planck-scale physics and LSV is at the heart of the SME.
	So, in the framework of the SME, large numbers of new coefficients are introduced, which are constrained experimentally. Apart from these, it should be stressed on this point that SME is merely an effective field theory framework, meaning that it does not include all dynamics of other LV-specific theories such as the Einstein-aether theory \cite{Jacobson:2000xp} (for more details see \cite{Kostelecky:2003fs}).
	In the SME, matter-gravity couplings occur through some coefficients, leading to a breaking of the EEP \cite{Kostelecky:2010ze}. Actually, in the minimal SME, the purely gravitational part of Einstein-Hilbert action gets modified by employing new terms that induce LV corrections in the gravitational sector. Based on the fact that Lorentz symmetry is the foundation of both GTR and the standard model of particle physics, in experimental searches for LSVs one can take the  advantage of either gravitational or non-gravitational forces or both.

	In this paper, we are interested in a SME-based probing LSV via a novel gravitational compact object, the Sgr A*'s shadow. Recalling some well-known gravitational frameworks that in the last years have been devoted to constraining the Lorentz violating coefficients (LVCs) ${\bar s}^{\mu\nu}$ is useful. These 
	LVCs, in essence, come from the gravitational sector of the minimal SME framework and contain a squared matrix $4 \times 4$ with sixteen coefficients made by the temporal and spatial components $T$, and $X, Y, Z$, respectively. Because the matrix ${\bar s}^{\mu\nu}$ has two properties, symmetric and traceless, thereby, there are nine independent coefficients. Although we here briefly list some of the most famous ones, to find more details, one can see the review paper \cite{Hees:2016lyw}.

	\begin{itemize}
		
		\item {\bf  Cerenkov radiation} \cite{Kostelecky:2015dpa} (see also review paper \cite{Schreck:2018qlz}) -
		If a particle moves with a velocity exceeding the phase velocity of gravity, the gravitational Cerenkov radiation arises.  In this way, the relevant particle emits gravitational radiation until it loses enough energy to drop below the gravity speed. High-energy cosmic rays, in case of not losing whole their energy via Cerenkov radiation, can provide constraints on LVCs in the SME framework 
		\begin{align*}  \nonumber
			&	-1.9\times 10^{-13} < {\bar s}^{XX}+{\bar s}^{YY}-2 {\bar s}^{ZZ} < 1.3 \times 10^{-13}\,, \quad 
			-3.9 \times 10^{-14}< {\bar s}^{XY},\quad {\bar s}^{YZ}< 6.2 \times 10^{-14} \,, \nonumber \\
			&	-5.4 \times 10^{-14} < {\bar s}^{XZ}< 5.4 \times 10^{-14} \,, \quad 
			2.8 \times 10^{-14}< {\bar s}^{TX}< 2.8 \times 10^{-14} \,, \nonumber \\
			&	-3.1 \times 10^{-14} < {\bar s}^{TY}< 2.4 \times 10^{-14} \,, \quad 
			1.7 \times 10^{-14}< {\bar s}^{TZ}< 2.4 \times 10^{-14} \,. \nonumber\\
			&	{\bar s}^{TT} > - 6\times 10^{-15}\,, \quad -9\times 10^{-14}< {\bar s}^{XX}-{\bar s}^{YY} < 1.2 \times 10^{-13}\,.
		\end{align*}	
		\item {\bf Lunar Laser Ranging (LLR)} \cite{Battat:2007uh,Bourgoin:2016ynf} - LLR experiment allows high precision measurements of the light travel time ranging from short laser pulses emitted by an LLR station until getting it again at a receiver station upon reflecting it by a lunar retro-reflector. From this experiment one gets the following constraints for LVCs 
		\begin{align*}\label{sLLR} \nonumber
			&{\bar s}^{XX}-{\bar s}^{YY} = (0.6\pm 4.2)\times 10^{-11}\,, \quad {\bar s}^{XY}=(-5.7\pm 7.7)\times 10^{-12}\,, \quad
			{\bar s}^{XZ} = (-2.2\pm 5.9)\times 10^{-12} \,, \\
			&{\bar s}^{TZ} = (-0.9\pm 1.0)\times 10^{-8}\,, \quad \quad 		{\bar s}^{TY}+ 0.43 {\bar s}^{TZ}=(6.2\pm 7.0)\times 10^{-9}\,,\quad
			{\bar s}^{XX} + {\bar s}^{YY}-4.5 {\bar s}^{ZZ} = (2.3\pm 4.5)\times 10^{-11}\,. \nonumber
		\end{align*}
		\item {\bf Atomic gravimetry} \cite{Muller:2007es,Chung:2009rm} - 
		Gravimeter tests are the most sensitive Earth-based experiments looking for LSV in the minimal gravity sector of SME.  Owing to Earth rotation, the signal recorded in a gravimeter would be modulated by the LSV terms, resulting in the following constraints
		\begin{align*}  
			&{\bar s}^{XX} - {\bar s}^{YY} =(4.4\pm 1.1)\times 10^{-9}\,, \quad {\bar s}^{XY}=(0.2\pm 3.9)\times 10^{-9}\,, \quad {\bar s}^{XZ}=(-2.6\pm 4.4)\times 10^{-9}\,, \\
			&{\bar s}^{YZ}=(-0.3\pm 4.5)\times 10^{-5}\,, \quad \quad \quad
			{\bar s}^{TX}=(-3.1\pm 5.1)\times 10^{-5}\,,~~ 
			{\bar s}^{TY}=(0.1\pm 5.4)\times 10^{-5}\,,\\ 
			&{\bar s}^{TZ}=(1.4\pm 6.6)\times 10^{-9}\,.  \nonumber
		\end{align*}
		
		\item {\bf Binary pulsars timing} \cite{Shao:2014oha,Shao:2014bfa}
		- 
		After observing the first binary star system composed of a neutron star and a pulsar known as PSR 1913+16 \cite{Hulse}, the binary pulsar systems have become a laboratory for measuring or constraining free parameters of gravitational theories. Regarding the SME coefficients one has   
		\begin{eqnarray}\label{sBinarySystem}\nonumber
			{\bar s}^{TX}&=&(0.05\pm 5.25)\times 10^{-9}\,, \quad 
			{\bar s}^{TY}=(0.5\pm 8.0)\times 10^{-9}\,, \quad
			{\bar s}^{TZ}=(-0.05\pm 5.85)\times 10^{-9}\, \nonumber \\
			|{\bar s}^{TT}| &<& 2.8 \times 10^{-4}\,, \quad {\bar s}^{XX}-{\bar s}^{YY}=(0.2\pm 9.9)\times 10^{-11}\,, \quad  
			{\bar s}^{XX}+{\bar s}^{YY}-2{\bar s}^{ZZ}=(-0.05\pm 12.25)\times 10^{-11}\,, \nonumber \\
			{\bar s}^{XY}&=&(0.05\pm 3.55)\times 10^{-11}\,, \quad 
			{\bar s}^{XZ}=(0.0\pm 2.0)\times 10^{-11} \,, \quad 
			{\bar s}^{YZ}=(0.0\pm 3.3)\times 10^{-11} \,, \nonumber 
		\end{eqnarray}
		
		\item {\bf Planetary ephemerides} \cite{Iorio:2012gr,Hees:2015mga} -
		Historically, the advance of the Mercurio perihelion rotating around the Sun was the first verification of GTR predictions. Since then, planetary ephemerides, are considered a robust tool to test GTR and constrain the modified theories of gravity. There are different effects arising from the LSV in the gravity sector of SME, that can have implications on planetary ephemerides analysis: effects on the light propagation and the orbital dynamics. From these one can infer the following bounds
		\begin{eqnarray}\label{sEphem}\nonumber
			{\bar s}^{XX} &-& {\bar s}^{YY}=(-0.8\pm 2.0)\times 10^{-10}\,, \quad 
			{\bar s}^{XX} + {\bar s}^{YY}-2{\bar s}^{ZZ}=(-0.8\pm 2.7)\times 10^{-10}\,, \\
			{\bar s}^{XY}&=&(-0.3\pm 1.1)\times 10^{-10}\,, \quad
			{\bar s}^{XZ}=(-1.0\pm 3.5)\times 10^{-11}\,, \quad
			{\bar s}^{YZ}=(8.8\pm 5.2)\times 10^{-12}\,, \quad \nonumber \\
			{\bar s}^{TX}&=&(-2.9\pm 1.1)\times 10^{-9}\,, \quad
			{\bar s}^{TY}=(0.3\pm 1.4)\times 10^{-8}\,, \quad
			{\bar s}^{YZ}=(-0.2\pm 5.0)\times 10^{-8}\,, \nonumber
		\end{eqnarray}
		
		\item {\bf Very Long Baseline Interferometry (VLBI)}  \cite{LePoncin-Lafitte:2016ocy}
		- VLBI is a type of astronomical interferometry employed in radio astronomy to measure the time difference in the arrival of a radio wavefront. The latter, in essence, is emitted by a far quasar and is measured through at least two Earth-based radio telescopes. VLBI provides the possibility of probing the gravitational sector of the minimal SME framework via constraining the time-time coefficient ${\bar s}^{TT}= (-5\pm 8)\times 10^{-5}$.
		
		\item {\bf Gravity Probe B (GPB)} \cite{Bailey:2013oda,Overduin:2013kja}
		- Based on the prediction of GTR, a gyroscope due to orbiting move around a rotating body experiences two relativistic precessions relative to a far inertial frame: a geodetic drift in the orbital plane and a frame-dragging. The former is due to the motion of the gyroscope in the curved spacetime, and the latter is generated by the spin of the central body. By shedding light from a satellite-based experiment known as GPB on the evolution of the spin of a gyroscope, at the parameterized post-Newtonian (PPN) approximation, one infers the following bounds for the combinations of LVCs
		\begin{eqnarray} \label{sGPB}\nonumber
			{\bar s}^{TT}&+&970 ({\bar s}^{XX}-{\bar s}^{YY})-0.05 ({\bar s}^{XX}+{\bar s}^{YY}-2{\bar s}^{ZZ})+2895\, {\bar s}^{XY}-3235 {\bar s}^{XZ}-11240 {\bar s}^{YZ}=(0.7\pm 3.1)\times 10^{-3}\,, \nonumber \\
			{\bar s}^{XX}&-&{\bar s}^{YY}+3.025 {\bar s}^{XY}+1.05 {\bar s}^{YZ}=(-1.1\pm 3.8)\times 10^{-7}\, \nonumber.  
		\end{eqnarray}
	\end{itemize} 
	The above upper bounds, in essence, tell us about the different sensitivity levels of SME-based search for LSV, typically at the level of $10^{-5}$ to $10^{-15}$.  With the advent of gravitational waves, we also deal with a novel observational platform to provide a clean test of the Lorentz invariance in the pure-gravity sector of the minimal SME \cite{Kostelecky:2016kfm,Mewes:2019dhj,LIGOScientific:2017zic}. Although we are interested in searching for LSV within the gravity sector, LSV is not limited merely to that rather, it could occur in other sectors such as the electron, and the photon, too (see Ref. \cite{Kostelecky:2008bfz}). Concerning the non-minimal SME,  one can also find a number of constraints in Ref. \cite{Tasson:2016xib}. 
	
	In this regard, in this paper we will introduce another treasure of nature so-called shadow image of Sgr A*, as a novel framework for exploring LSV in gravity. It would be interesting to mention that despite some works \cite{Bonder:2021gjo,Bailey:2021jqb}, there are no exact black hole solutions for the general SME framework. As a subclass of minimal SME involving vector field, one can be mentioned to Bumblebee gravity-based Schwarzschild-like solution  \cite{Casana:2017jkc}. It would be interesting to note that the modified Schwarzschild spacetime solution released in \cite{Casana:2017jkc} comes from  only a non-zero radial component, while in a recent paper \cite{Xu:2022frb} calculations are performed for the case of a non-zero temporal component of the bumblebee field.	
	Throughout this paper we will in fact follow our aim by comparing the minimal SME-inspired Schwarzschild metric obtained from a PPN approach in the weak-field limit, with the first image of Sgr A*. Generally, any extended framework of gravitation can yield corrections to the Newtonian potential, which in the PPN formalism is related to the metric tensor and, in this way, gives rise to a corrected Schwarzschild solution \cite{Will}. Apart from simplicity, the main reason for adopting the metric with spherical symmetry is related to the fact that there is no consensus on the value of Sgr A* spin and observation angle.
	Even though it seems that to describe the Sgr A*'s shadow, there is a consistency between large spin and low observation angle. Its opposite case, i.e., low spin and large observation angle, conclusively has not been ruled out yet \cite{EventHorizonTelescope:2022xnr,EventHorizonTelescope:2022urf}.
	In such a position, we prefer to adopt a conservative approach and, in agreement with the philosophy of simplicity, take into account a simple metric corrected by LV terms enjoying spherical symmetry. 
	
	The outline of the rest of this manuscript is as follows. In the light of the effective Newtonian potential obtained from the PPN approximation within the framework of the pure-gravity sector of the minimal SME, in section \ref{secs.LV} we derive a Schwarzschild metric corrected by the LVCs. In the section \ref{secs.Shadow}, we extract shadows related to the Schwarzschild metric corrected by LV terms. In section \ref{secs.sgr} we then confront the resulting shadows to the shadow of Sgr A* recently inferred by the EHT's team, to set some constraints on the LVCs induced by the underlying minimal SME. Finally, we present a summary of our discussions and conclusions in section \ref{secs.con}. Throughout this paper, for simplicity, we work with the units $c=\hbar=1$.

	\section{SME inspired Schwarzschild metric}
	\label{secs.LV}
	By taking the minimal version of the SME in the Riemann spacetime limit, one deals with an effective action given by 
	\begin{equation}\label{GS}
		S=S_{HE}+S_m+S_{LV}\,,
	\end{equation}
	where $S_{HE}=(16\pi G)^{-1}\int d^4x e(R-2\Lambda)$ (here $e\equiv\sqrt{-g}$ is the determinant of the vierbein) is the standard Hilbert-Einstein action including the cosmological constant $\Lambda$, $S_m$ is the general matter action, and $S_{LV}$ \cite{Bailey:2006fd}
	\begin{equation}
		\label{LSaction}
		S_{LV}=\frac{1}{16\pi G}\int d^4x e\left(-uR+s^{\mu\nu}R^T_{\mu\nu}+t^{\kappa\lambda\mu\nu}
		C_{\kappa\lambda\mu\nu}\right)\,,
	\end{equation}
	with the trace-free Ricci  tensor $R^T$ and the Weyl conformal tensor $C_{\kappa\lambda\mu\nu}$, which address the LV gravitational couplings. Given that we are just interested in incorporating the gravitational interaction in SME,  we do not pay attention to action $S_m$, which includes LV matter-gravity coupling \footnote{In the light of probing the LSV via the MICROSCOPE space mission, it has been extracted some tight constraints for the matter-gravity couplings \cite{Pihan-LeBars:2019qsd} (see also \cite{Mo:2019pil}).}. The coefficients $u$, $s^{\mu\nu}$ and $t^{\kappa\lambda\mu\nu}$ are real and dimensionless so that the second and third ones satisfy the Ricci and Riemann properties (symmetry with respect to switching indices and also obeying the Bianchi identity), respectively, and are also traceless, i.e., $s^{\mu}_{{\phantom \mu} \mu}=0\,, \quad
	t^{\kappa\lambda}_{\phantom{\kappa\lambda}\kappa\lambda}=0\,, \quad
	t^{\kappa}_{\phantom{\kappa}\mu\kappa\lambda}=0$. By restricting ourselves to the case $u=0$ and $t^{\kappa\lambda\mu\nu}=0$, thereby,  across this paper, the origination of Lorentz violation degrees of freedom come just from the coefficients $s^{\mu\nu}$ \footnote{To avoid a challenge with the Bianchi identity, in Riemann's geometry is assumed the LVCs are dynamical fields, and the LV owing to a spontaneous symmetry breaking, is caused \cite{Kostelecky:2003fs,Bluhm:2004ep,Bluhm:2007bd}. The vacuum expectation value of LV fields may be non-zero. By adopting a linearized gravity limit, one can be integrated out the fluctuations surrounding the vacuum values, meaning that just the vacuum expectation values of the SME's coefficients stay, and they may affect observations \cite{Bailey:2006fd}. The coefficient $\bar u$, in the minimal SME, is unobservable since it is nothing but a rescaling of the gravitational constant, while the coefficients $\bar t^{\kappa\lambda\mu\nu}$ do not play, at the PPN level, any role (t-puzzle) \cite{Bonder:2015maa}. It would be interesting to note that authors recently in Refs \cite{Reyes:2022mvm,Reyes:2021cpx} have developed a Hamiltonian formalism (ADM-decomposition) for the background fields $u$ and $s^{\mu\nu}$ in the minimal gravitational SME. By employing these background fields also has been proposed the diffeomorphism violation-based cosmology \cite{Reyes:2022dil,ONeal-Ault:2020ebv}.}. Now, by varying the action $S$ with respect to the background metric, we have the following extended Einstein field equations 
	\cite{Bailey:2006fd}
	\begin{equation}\label{fieldequations}
		G^{\mu\nu}-(T^{Rs})^{\mu\nu}=8\pi G\, T_g^{\mu\nu}\,,
	\end{equation}
	where 
	\be\label{TRs}
	(T^{Rs})^{\mu\nu} = \frac{1}{2}g^{\mu\nu}(R_{\alpha\beta}-
	\nabla_\alpha \nabla_\beta) s^{\alpha\beta} 
	+  \frac{1}{2}\left(\nabla_\alpha \nabla^\mu s^{\alpha\nu}+\nabla_\alpha \nabla^\nu s^{\alpha\mu} -\nabla_\alpha \nabla^\alpha s^{\mu\nu}\right)\,. 
	\ee
	Here $G^{\mu\nu}=R^{\mu\nu}-(R/2)g^{\mu\nu}$ is the standard
	Einstein tensor. Avoiding details, the PPN approximation studied in \cite{Bailey:2006fd} for the theory at hand yields a two-point-particle Lagrangian, which gives the effective equations of motion for a two bodies system (with masses $M$ (heaviest) and test particle $m$), in the coordinate acceleration. If $M \gg m$ so that the heaviest body $M$ be at rest relative to $m$, then the effective two bodies Lagrangian in the presence of Lorentz violation coefficients $\bar{s}^{\mu\nu}$,  takes the following form 
	\be
	L = \frac{1}{2}m v^2 + \frac{G m M}{r}\left(1 + \frac{3}{2}\bar{s}^{00} +
	\frac{1}{2} \bar{s}^{jk}\frac{x_j x_k}{r^2}\right) 
	- \frac{G m M}{r}\left(3 \bar{s}^{0j}v_j + \bar{s}^{0j}\frac{x_j}{r}v_k\frac{x_k}{r}\right)~,
	\label{KL}
	\ee
	where $00$ denotes the time-time ($TT$) coefficient and indexes $j, k$, run from 1-3 (equivalent to $x_1=X,x_2=Y$ and $x_3=Z$ in Cartesian coordinates) and $v^2=v_1^2 + v_2^2 + v_3^2$, $r^2 = x_1^2 + x_2^2 + x_3^2$, $v_k = \dot{x_k}$ (the derivative is taken with respect to the coordinate time). By adopting a stationary and weak gravitational field regime in which the test particle moves slowly in comparison with the speed of light ($v \ll 1$), one can then discard the terms depending on the velocity $v$ in (\ref{KL}). As a result, the effective potential reads
	\be
	V(r)=\frac{U(r)}{m}= -\frac{GM}{r}\left(1 + \frac{3}{2}\bar{s}^{00} +
	\frac{1}{2} \bar{s}^{jk}\frac{x_j x_k}{r^2}\right)~.
	\label{LV1}
	\ee
	However, this is not the final form since  the scalar factor
	$(1 + 3\bar{s}^{00}/2)$ merely rescales the gravitational constant and can be absorbed in it \cite{Bailey:2006fd}. So, the function $V(r)$ modified by Lorentz violation coefficients $\bar{s}^{\mu\nu}$ is given by the following expression 
	\be
	V(r) = -\frac{G_{\rm eff}M}{r}\left(1 +  \bar{s}^{jk}_{\rm eff}\frac{x_j x_k}{r^2}\right)~,
	\label{LV2}
	\ee
	where
	\be
	G_{\rm eff}=G\left(1 + \frac{3}{2}\bar{s}^{00}\right) \ , \quad \bar{s}^{jk}_{\rm eff}=\frac{\bar{s}^{jk}}{2+3\bar{s}^{00}}~.
	\ee
	It is important to note two points here. First, the term $\bar{s}^{jk}x_j x_k/r^2$ cannot be reabsorbed since in fact it is a term with dependency on the directions $x_j x_k/r^2$.
	Second, since the factors containing $\bar{s}^{00}$ are unobservable, by the end of this manuscript we shall work with $G_{\rm eff}\equiv G$ and $\bar{s}^{jk}_{\rm eff}\equiv \bar{s}^{jk}$.

	Now we  take the effective potential produced by a general metric of the form
	\be
	ds^2 = f(r)dt^2 - g_{ik}(x_1,x_2,x_3) dx^i dx^k~,
	\label{ik}
	\ee
	where $r=|\mathbf x|=\sqrt{x_1^2 + x_2^2 + x_3^2}$. As a spacial case, the above general metric produces the Schwarzschild metric in the standard form
	\be\label{SC}
	ds^2 = \left(1-\frac{2GM}{r}\right)dt^2 - \left(1-\frac{2GM}{r}\right)^{-1}dr^2 - r^2d\Omega^2~, 
	\ee 
	where $d\Omega^2=\sin^{2}\theta d\theta^2+d\varphi^2$. As is well known, any general metric of the form
	\be
	ds^2 = f(r)dt^2 - f(r)^{-1}dr^2 - g(r)d\Omega^2~,
	\label{gm}
	\ee
	can be brought into the form (\ref{ik}). In case of the metric be in the form (\ref{ik}), in Cartesian coordinates, by utilizing the well known procedures in Ref. \cite{Weinberg2008}, one can easily show that the effective Newtonian potential is of the form
	\be
	V(r) \ \simeq \ \frac{1}{2} \ (f(r)-1)~.
	\ee
	More exactly, the effective Newtonian potential is produced
	by the metric given  in (\ref{ik}) for  a point particle moving slowly in a stationary and weak gravitational field, which is equivalent to quasi-Minkowskian spacetime far way from the source, i.e., $r \to \infty$.
	As a consequence, the metric (\ref{gm}) is able to mimic the corrected Newtonian potential (\ref{LV2}) with the LV terms, where the laps function $f(r)$ is given by
	\be
	f(r) = 1 -\frac{2GM}{r}\big(1 + \bar{s}^{jk}\chi_{jk}(\theta, \phi)\big)~,
	\label{LVM}
	\ee
	where we introduced the standard spherical coordinates ${\bf x}=r (\sin\theta\cos\phi, \sin\theta\sin\phi, \cos\theta)$ and $x_j x_k / r^2 = \chi_{jk}(\theta, \phi)$.
	Although, in case of relaxing the LVCs i.e., $\bar{s}^{jk} \rightarrow 0$, the standard form of Schwarzschild is recovered, (\ref{LVM}) addresses merely a weak-field based black hole solution in the framework of the SME. Notice that, in the context of the scenario at hand explicit angular dependency $\chi_{jk}(\theta, \phi)$ will cause trouble. To bypass the angular dependency of $\chi_{jk}(\theta, \phi)$, we follow two ways.
	First, taking a overall averaging over $(\theta, \phi)$, i.e., $<x^jx^k>=\frac{r^2}{3}\delta^{jk}$, resulting in 
	\be
	f_{I}(r) = 1 -\frac{2GM}{r}\big(1 + \frac{\xi}{3}\big)~,~~~~~\xi=\bar{s}^{ii}
	\equiv\bar{s}^{XX}+\bar{s}^{YY}+\bar{s}^{ZZ}.
	\label{LVM2}
	\ee
	Second, by opening the expression $\bar{s}^{jk}\chi_{jk}(\theta, \phi)$ and putting $\theta=\pi/2$, we have 
	$(\bar{s}^{XX}-\bar{s}^{YY}) cos^2\phi+\bar{s}^{YY}+\bar{s}^{XY}sin 2\phi$. 
	By taking an average over $(\phi)$, the lapse function (\ref{LVM}), reads as
	\be
	f_{II} (r) = 1 -\frac{2GM}{r}\big(1 +\frac{\eta}{2} \big)~,~~~~\eta=\xi-\bar{s}^{ZZ}\equiv\bar{s}^{XX}+\bar{s}^{YY}.
	\label{LVM1}
	\ee
	Note that, setting $\theta=\pi/2$ is safe since in calculating spherically symmetric shadows, one commonly works in the equatorial plane. 
		In some special positions, one can restrict angular dependency in the metric (\ref{LVM}) to positive values and assume that $|\chi_{jk}(\theta, \phi)|\simeq 1$ \cite{Lambiase:2017adh,Blasone:2021phx}. For the distinction between these two, we labeled them with subscripts ''I'' and ''II''. Our main aim in the following is to constrain the dimensionless parameters $\xi$ and $\eta$ (including two different combinations of spatial diagonal LVCs) in the light of the first image released of Sgr A*'s shadow. Two comments are in order here. First, due to the spherical symmetry of the metric at hand, in this scenario we no longer are able to address the off-diagonal LVCs.
	Second, the tracelessness condition  $s^{\mu}_{{\phantom \mu} \mu}=0$ i.e., $\bar{s}^{TT}-\xi=0$ allows us to indirectly apply some constraints on the temporal diagonal LVC $\bar{s}^{TT}$, too. The upper limits, which in the following will be extracted for $\xi$ work for $\bar{s}^{TT}$, too. 
	
	Given that the LV-corrected lapse function (\ref{LVM}) has not been derived from an exact manner and merely is an approximate solution of the underlying minimal SME theory, one has to be careful in the use of the PPN formalism.
	Recall that in case of carrying higher-order corrections in the PPN approach, it can be a very effective framework for the description of the strong field regimes with the fast motion of particles \cite{Will:2011nz}. However, as it is obvious, in this scenario we restricted ourselves to leading corrections at the linear level of PPN. It is for that by going beyond the linear level, i.e., taking higher-order corrections induced by the SME into account in the PPN, there is no longer a guarantee that subsequently, the relevant lapse function be the solution of the theory at hand with equations (\ref{LSaction})-(\ref{TRs}).

	\begin{figure}[ht!]
		\centering
		\includegraphics[scale=0.6]{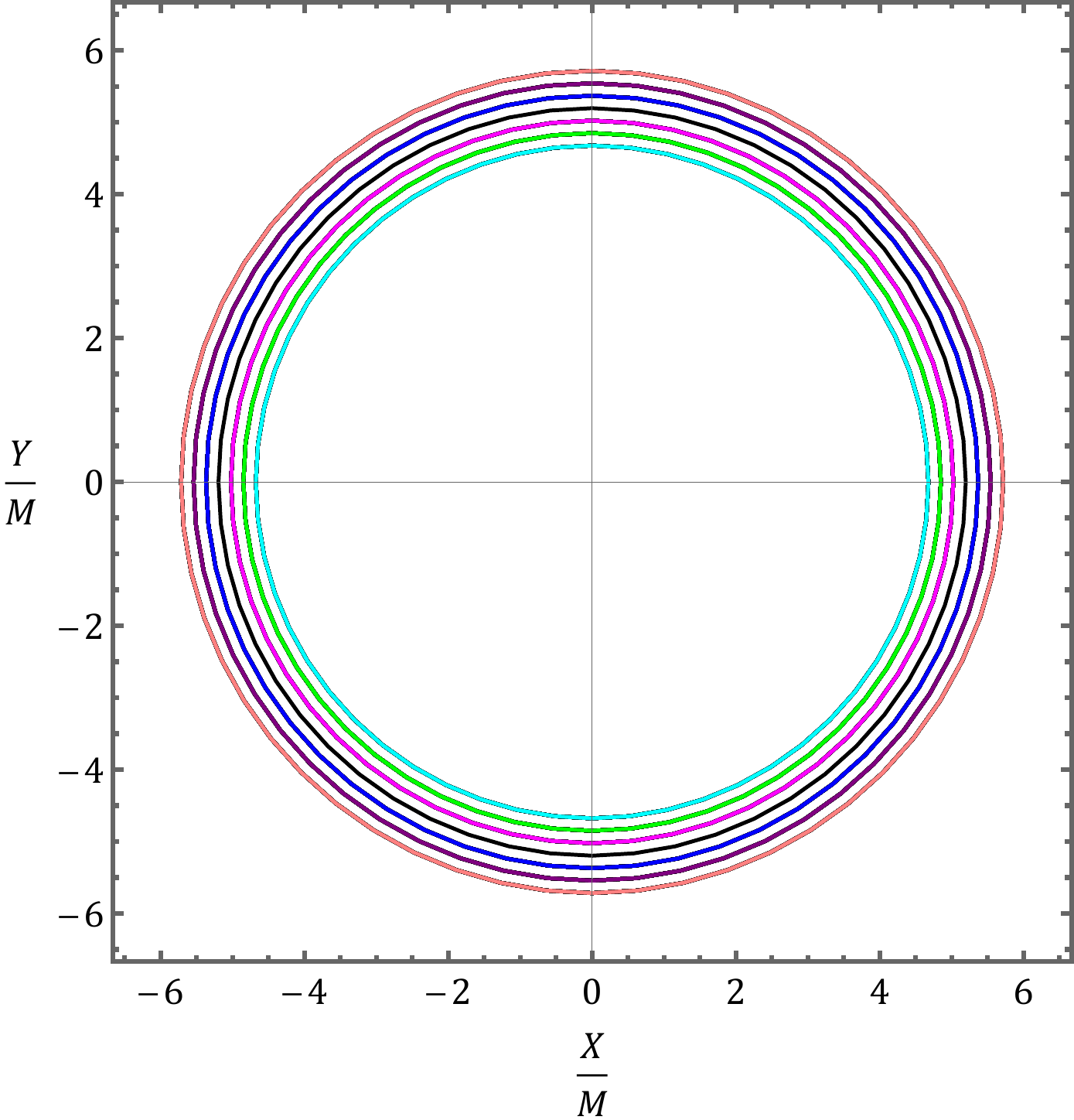}~~~~
		\includegraphics[scale=0.6]{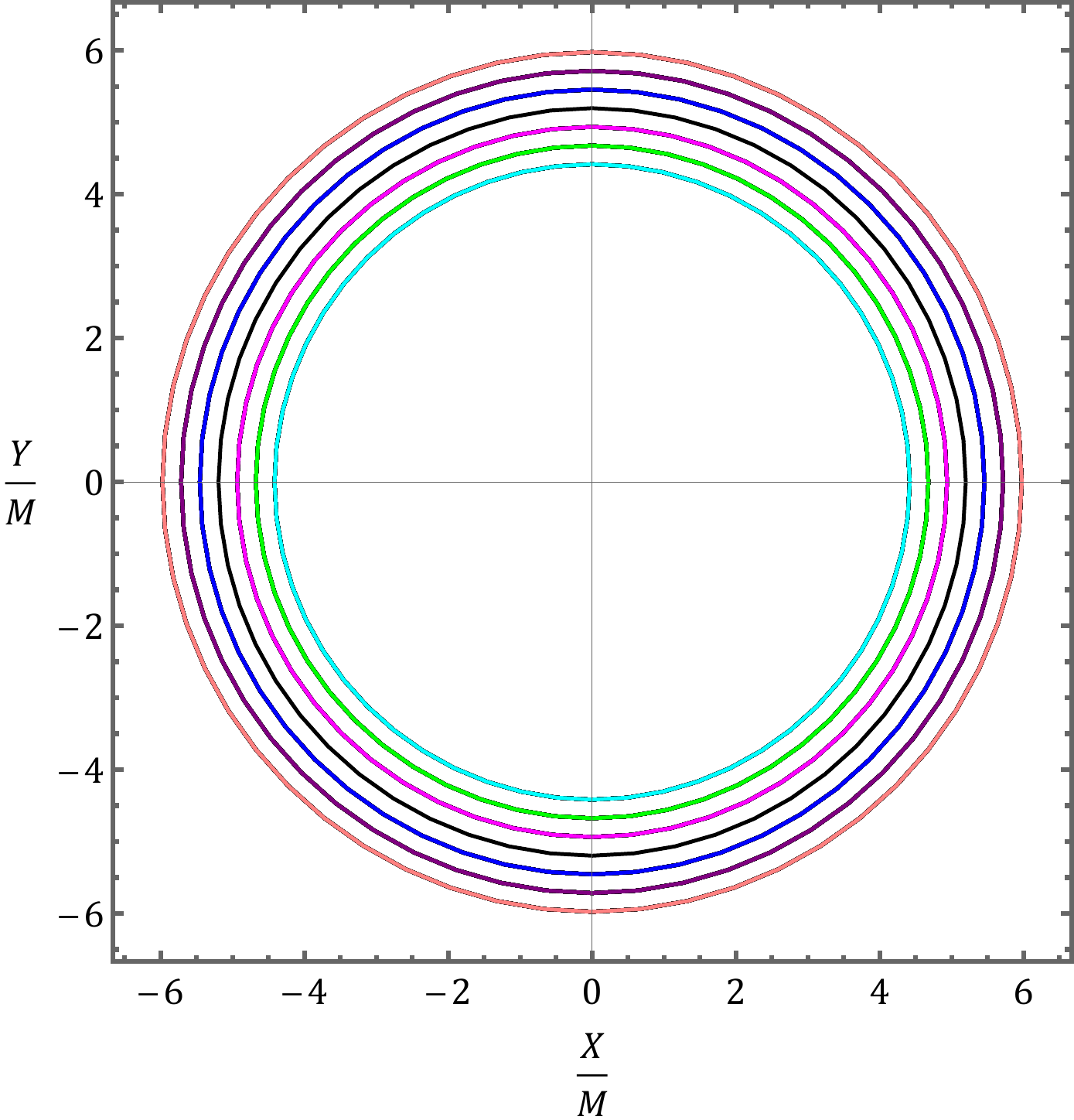}
		\caption{Apparent shadow of the Schwarzschild metric corrected by lapse functions: $f_I(r)$ (left) and $f_{II}(r)$ (right). In both panels values: $-0.3,-0.2,-0.1,0,0.1,0.2,0.3$ (from cyan curve to pink) fixed for the LVCs $\xi$ (left) and $\eta$ (right), respectively. }
		\label{sh}
	\end{figure}
	
	\begin{figure}[ht!]
		\centering
		\includegraphics[scale=0.5]{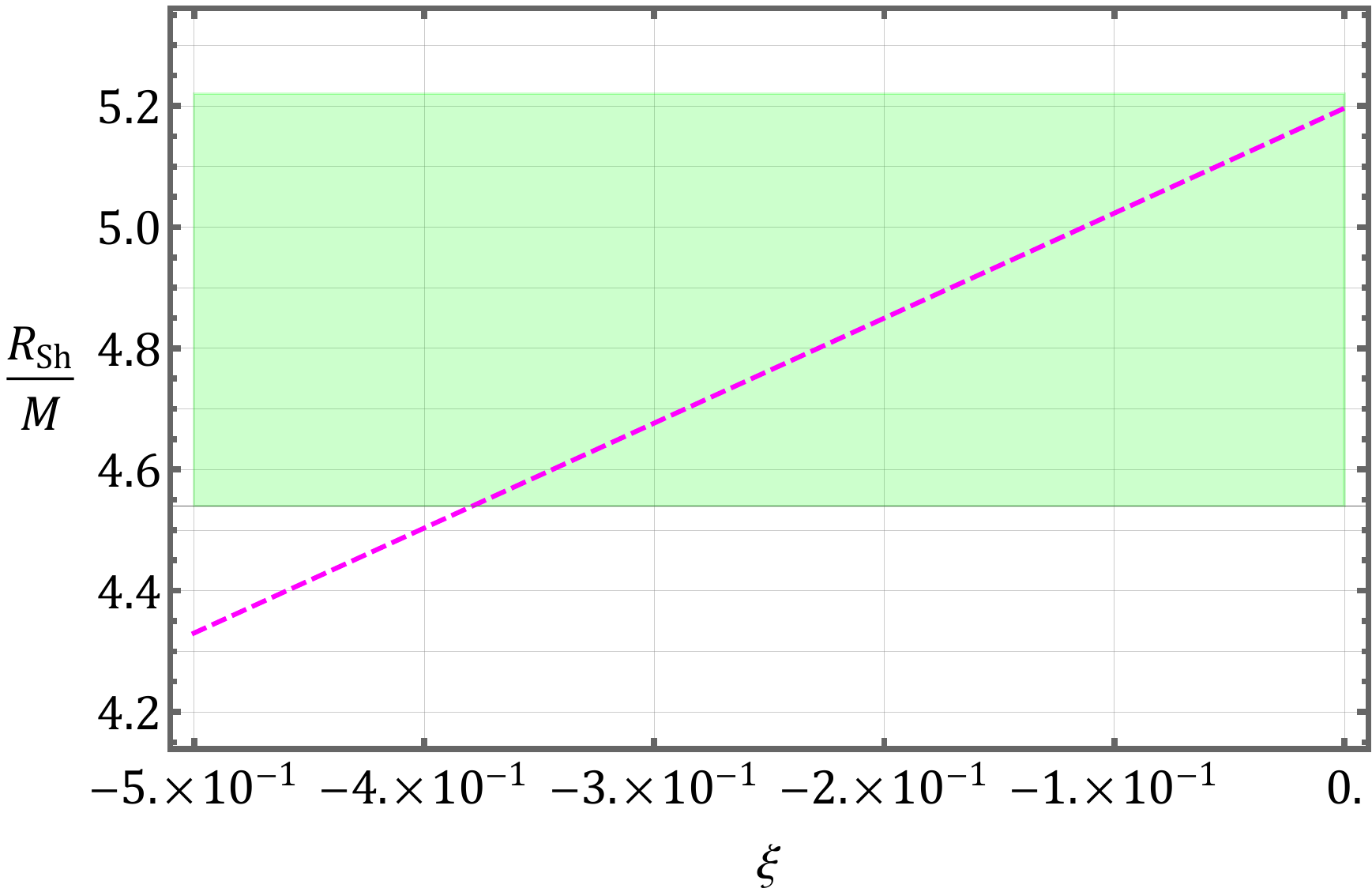}~~~~
		\includegraphics[scale=0.5]{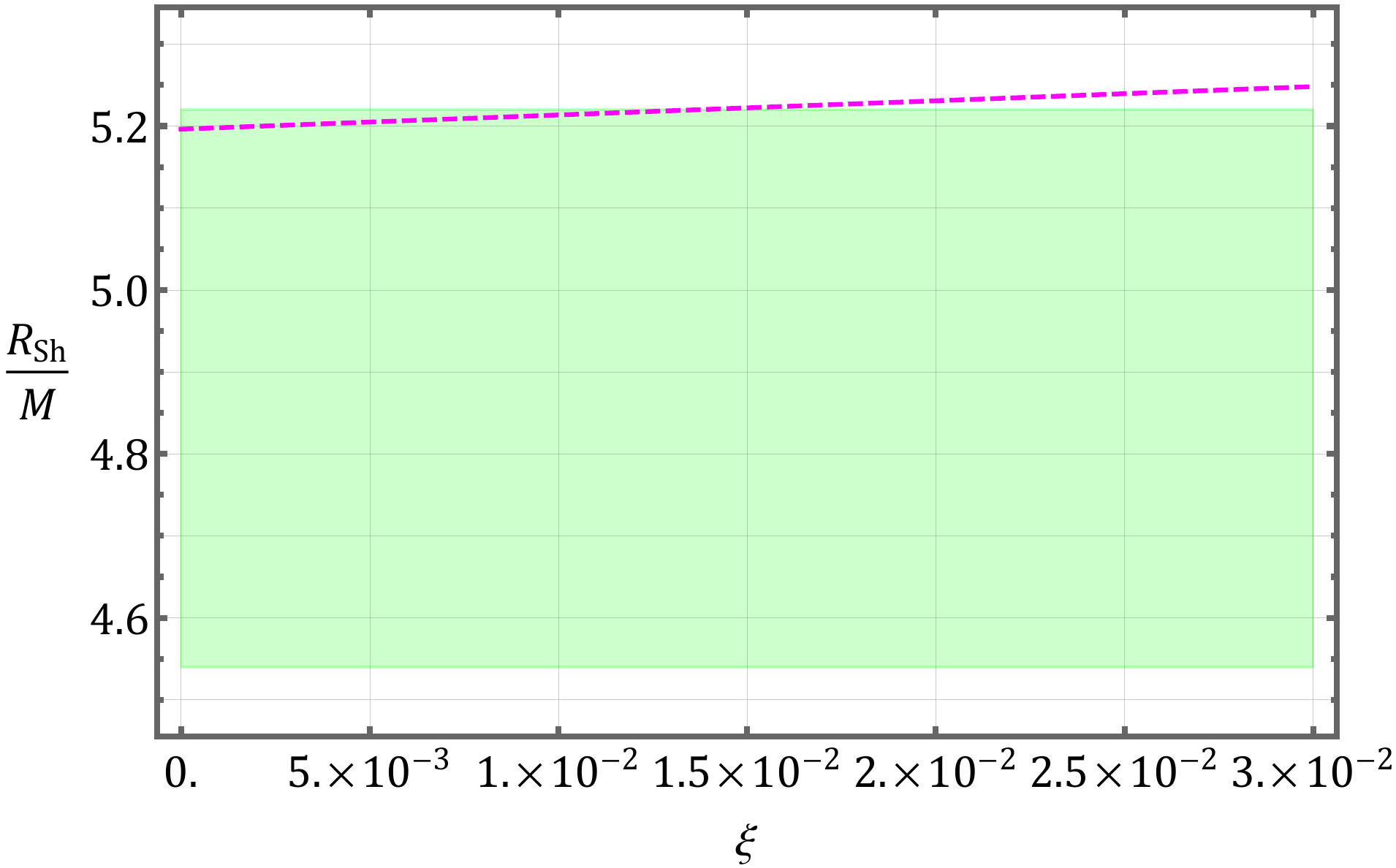}\\
		\includegraphics[scale=0.5]{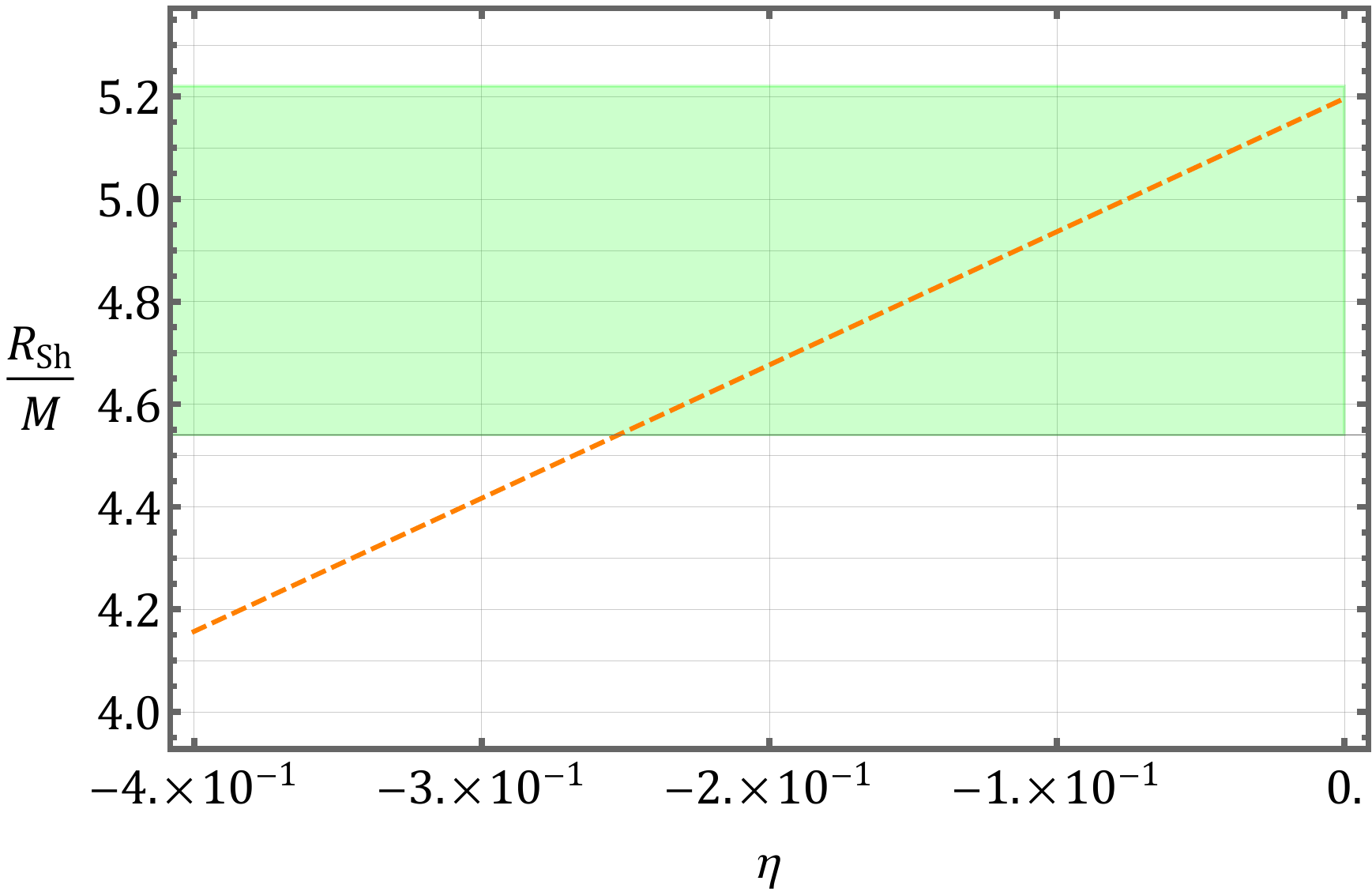}~~~~
		\includegraphics[scale=0.52]{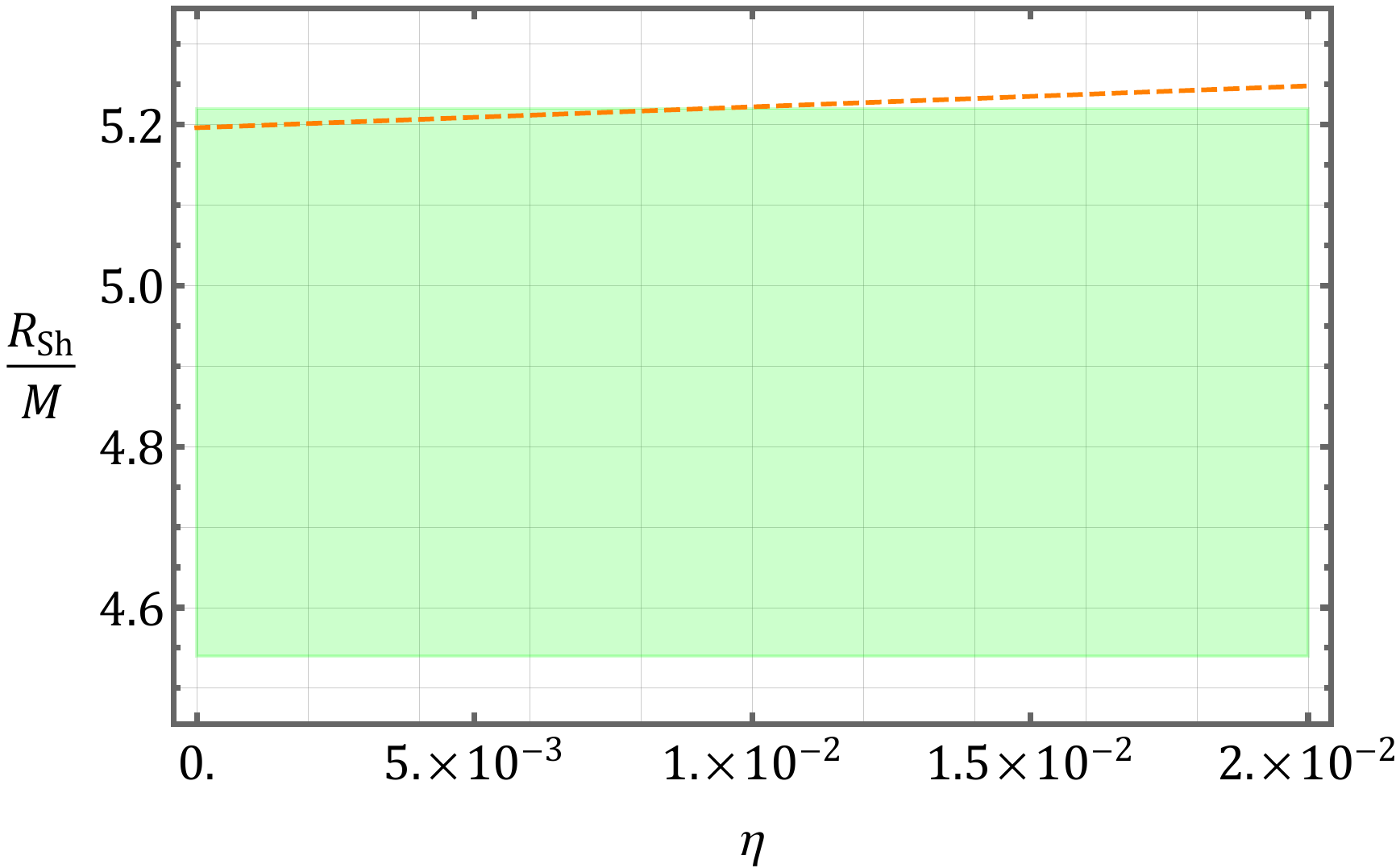}
		\caption{The predicted radius per unit mass $\frac{R_{sh}}{M}$ for the Schwarzschild metric corrected by lapse functions: $f_I(r)$ with the LVCs $\xi$ (up row) and $f_{II}(r)$ with the LVCs $\eta$ (bottom row). The shaded area mark the observationally determined radius per unit mass of Sgr A*'s shadow, namely $\frac{R_{\rm Sgr A*}}{M}$, within $1\sigma$ uncertainty. }
		\label{RR}
	\end{figure}

	\begin{figure}[ht!]
		\centering
		\includegraphics[scale=0.48]{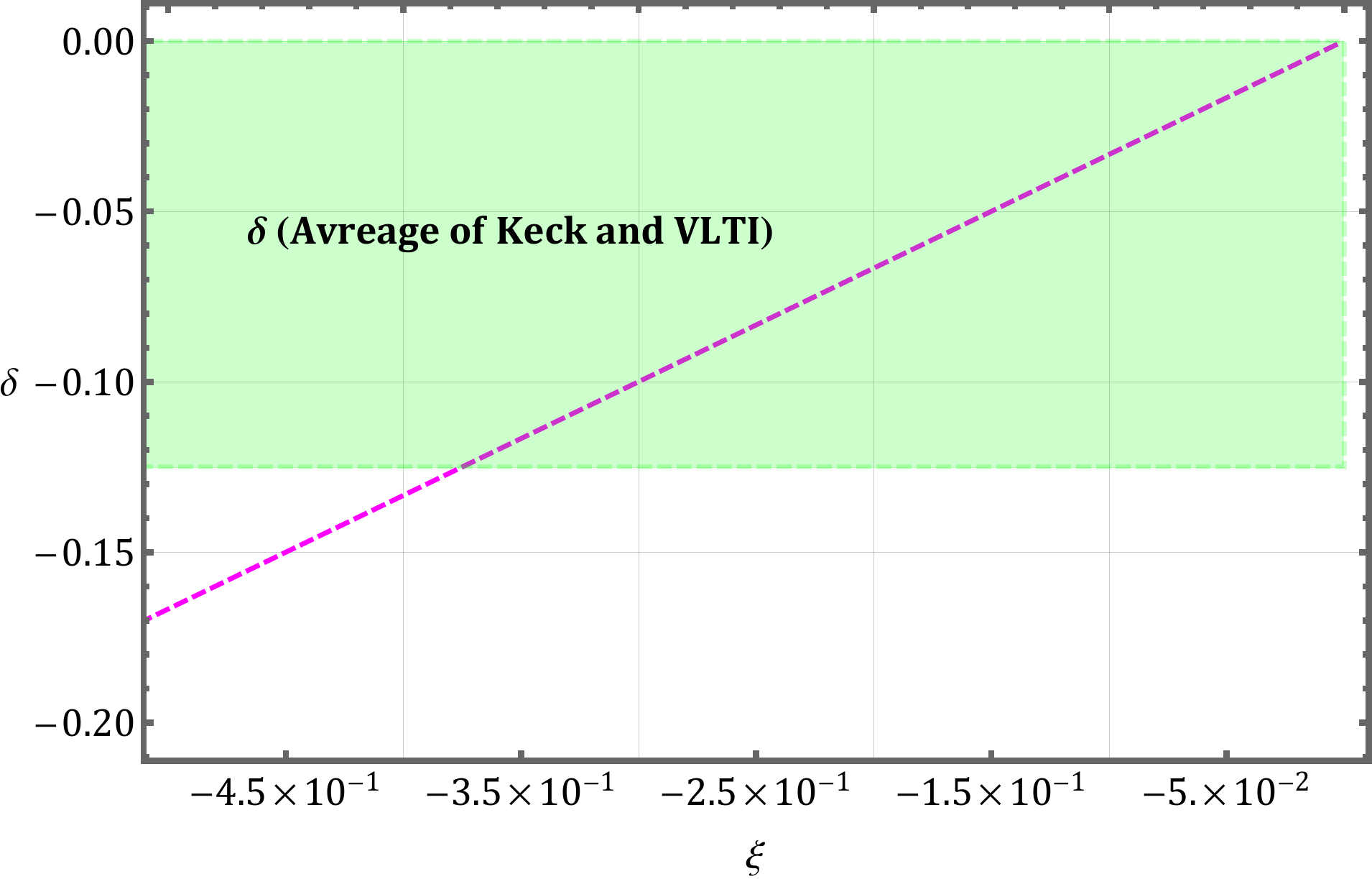}~~~~
		\includegraphics[scale=0.5]{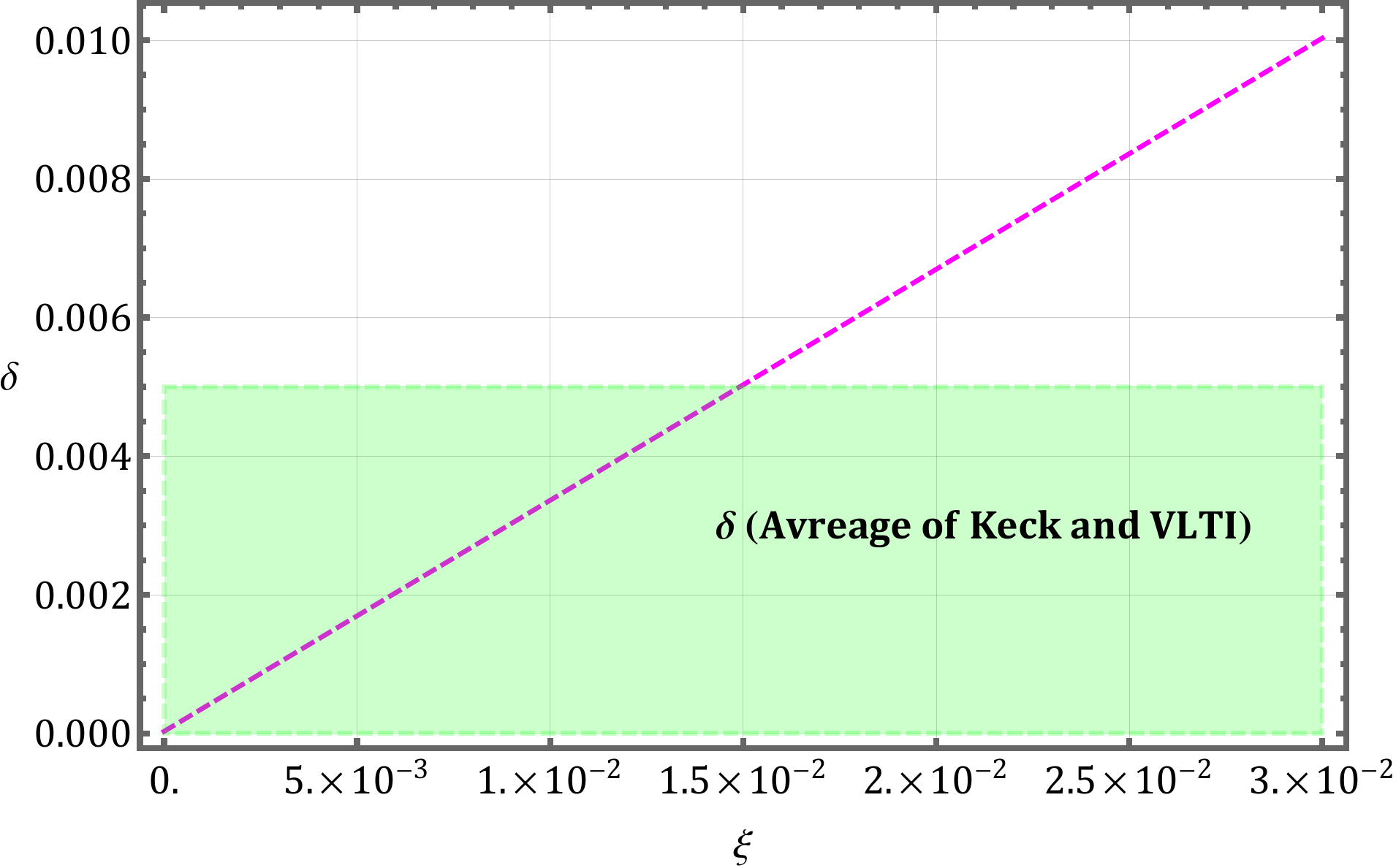}\\
		\includegraphics[scale=0.5]{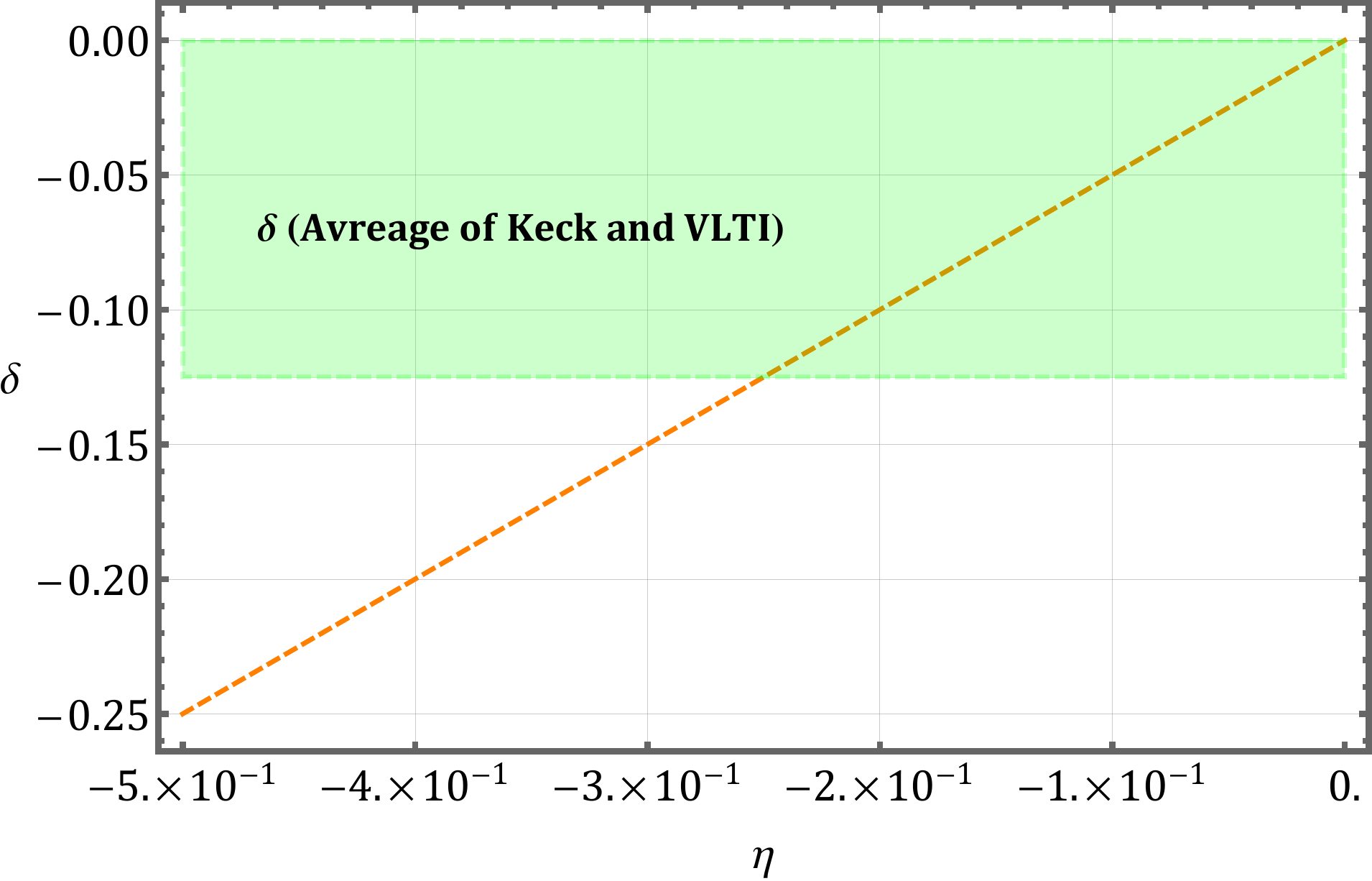}~~~~
		\includegraphics[scale=0.5]{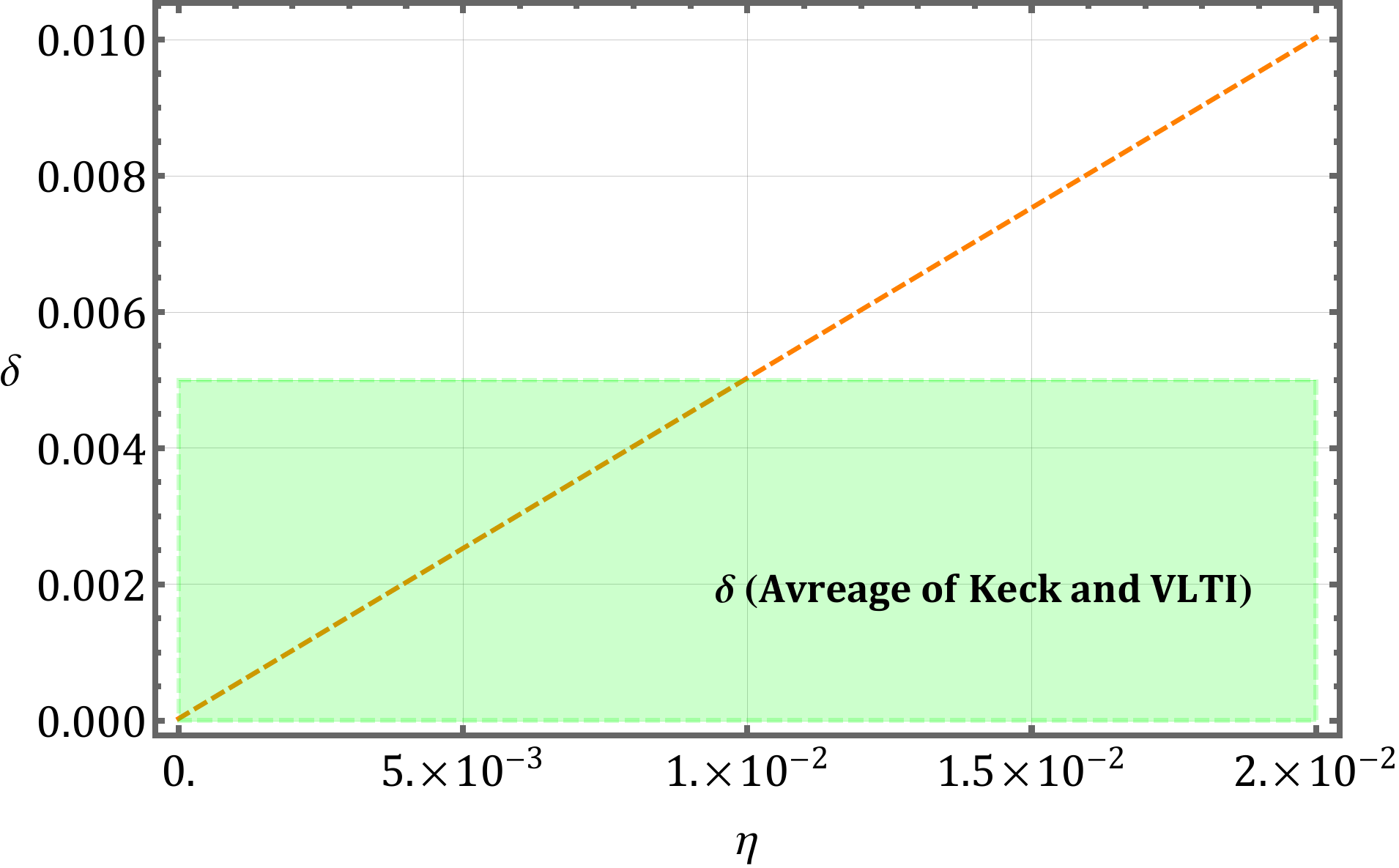}
		\caption{The predicted fractional deviation $\delta$ for the Schwarzschild metric corrected by lapse functions $f_I(r)$ with the LVCs $\xi$, and $f_{II}(r)$ with the LVCs $\eta$, respectively. The shaded area mark the observationally determined fractional deviations by estimation (\ref{eq:av}). To provide lower and upper bounds with enough resolution we have scanned values of negative and positive LVCs $\xi$ and $\eta$ within two complementary ranges: $-0.125\lesssim\delta<0$ (left panels) and $0<\delta\lesssim0.005$ (right panels), respectively.}
		\label{DDD}
	\end{figure}
	
	\begin{figure}[ht!]
		\centering
		\includegraphics[scale=0.48]{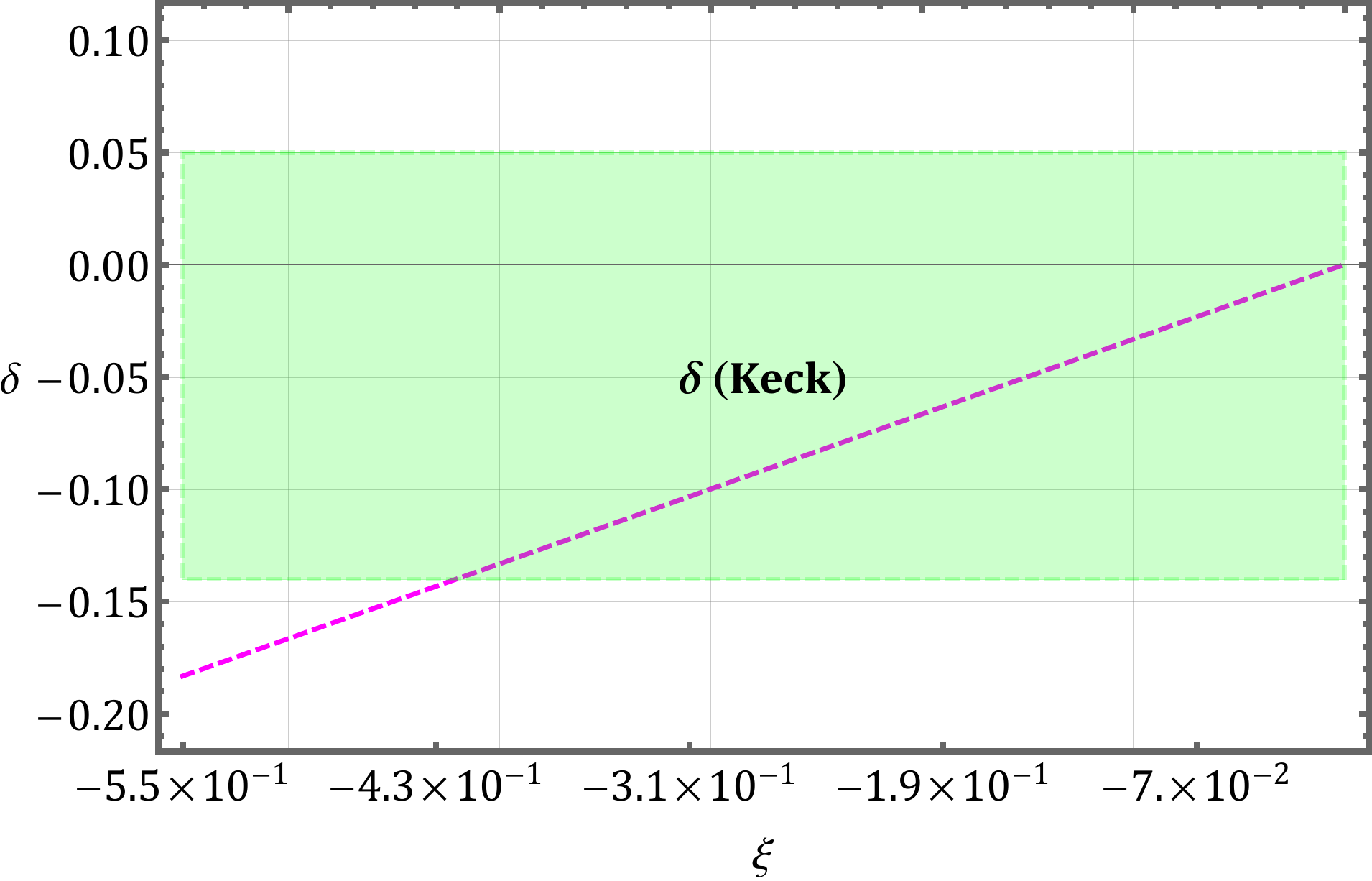}~~~~
		\includegraphics[scale=0.5]{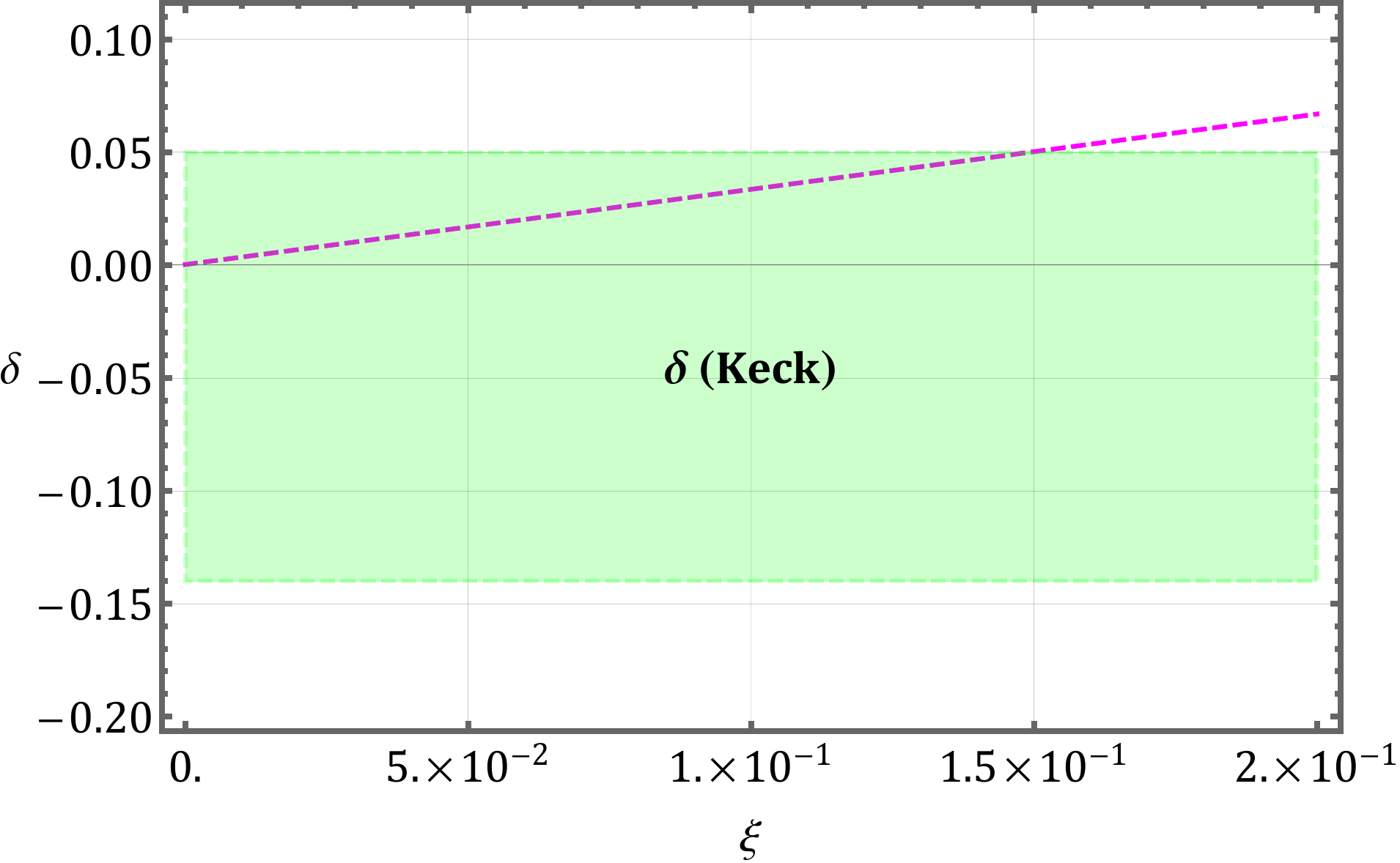}\\
		\includegraphics[scale=0.5]{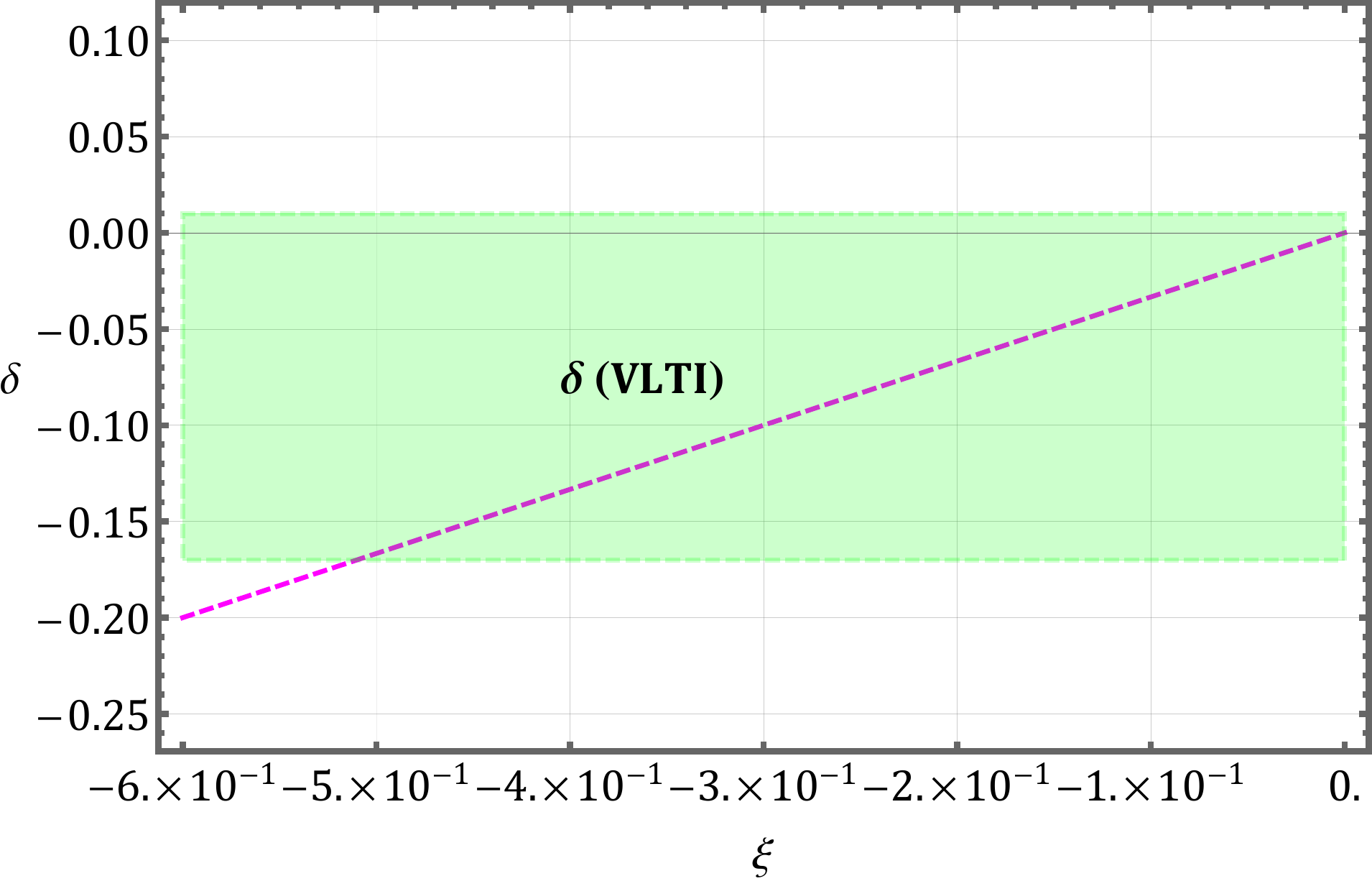}~~~~
		\includegraphics[scale=0.5]{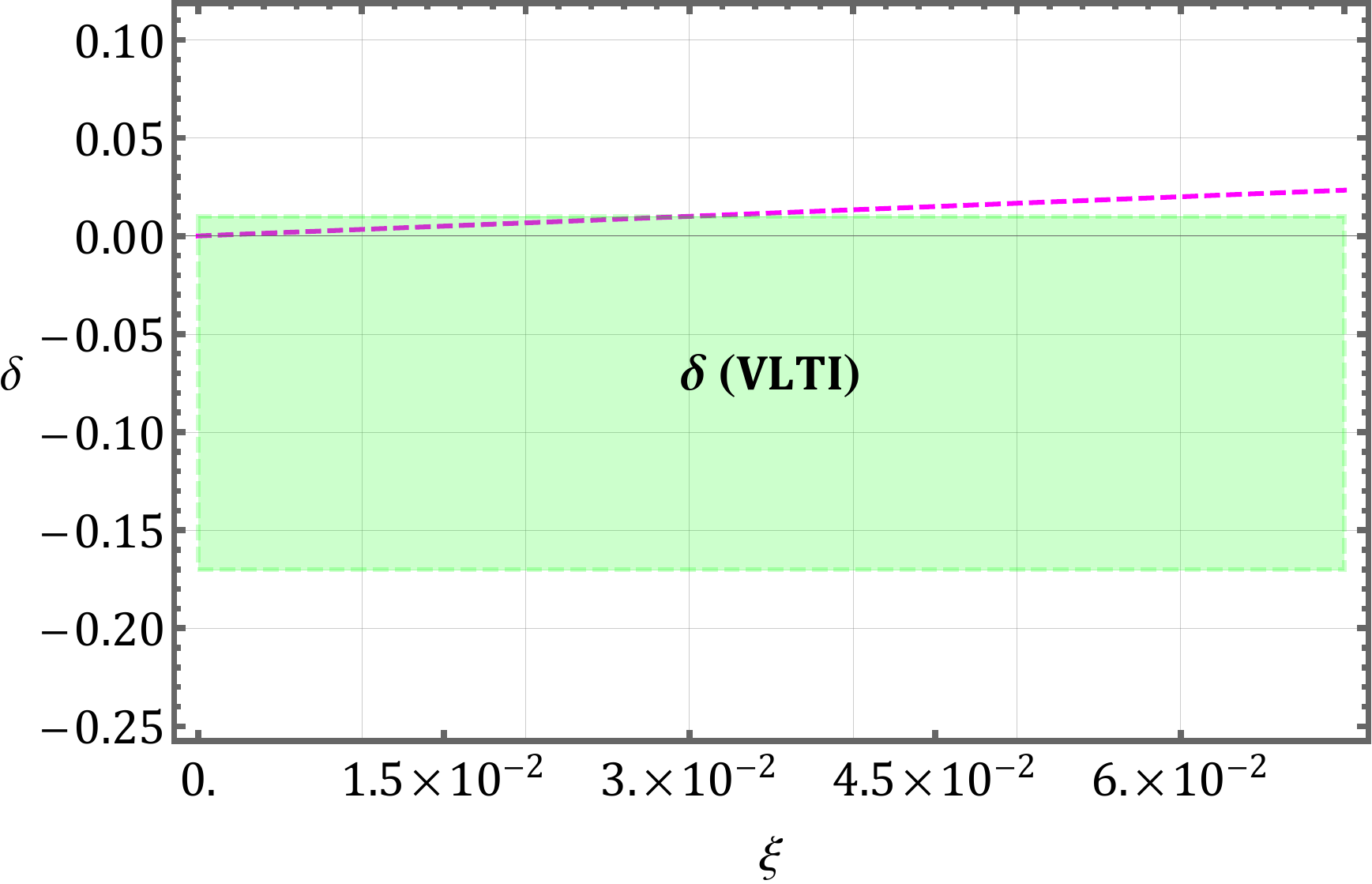}
		\caption{The predicted fractional deviation $\delta$ for the Schwarzschild metric corrected by lapse function $f_I(r)$ with the LVCs $\xi$. The shaded area in the up and bottom rows mark the observationally determined fractional deviations by Keck and VLTI measurements, respectively. }
		\label{DD1}
	\end{figure}

	\begin{figure}[ht!]
		\centering
		\includegraphics[scale=0.48]{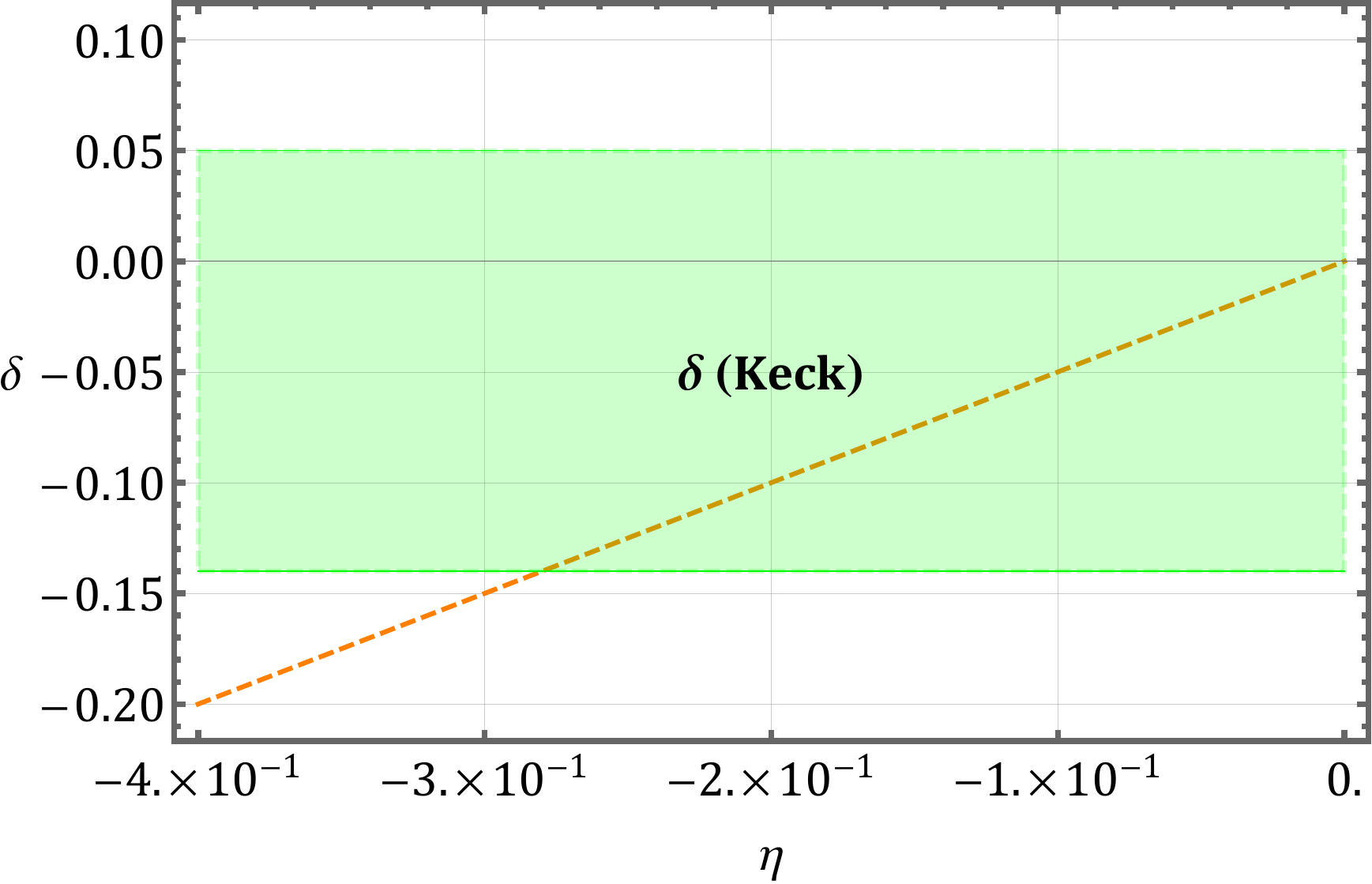}~~~~
		\includegraphics[scale=0.5]{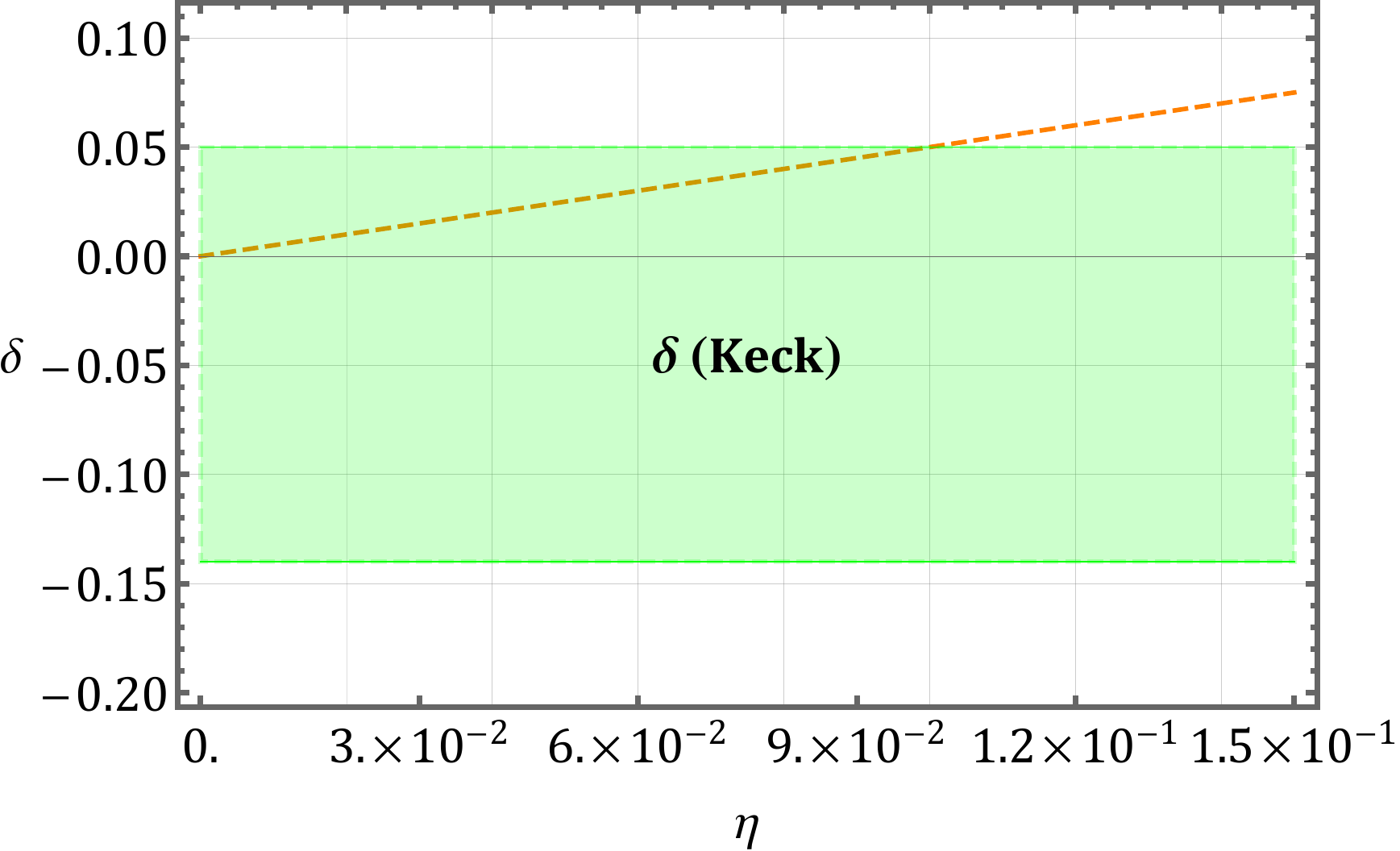}\\
		\includegraphics[scale=0.5]{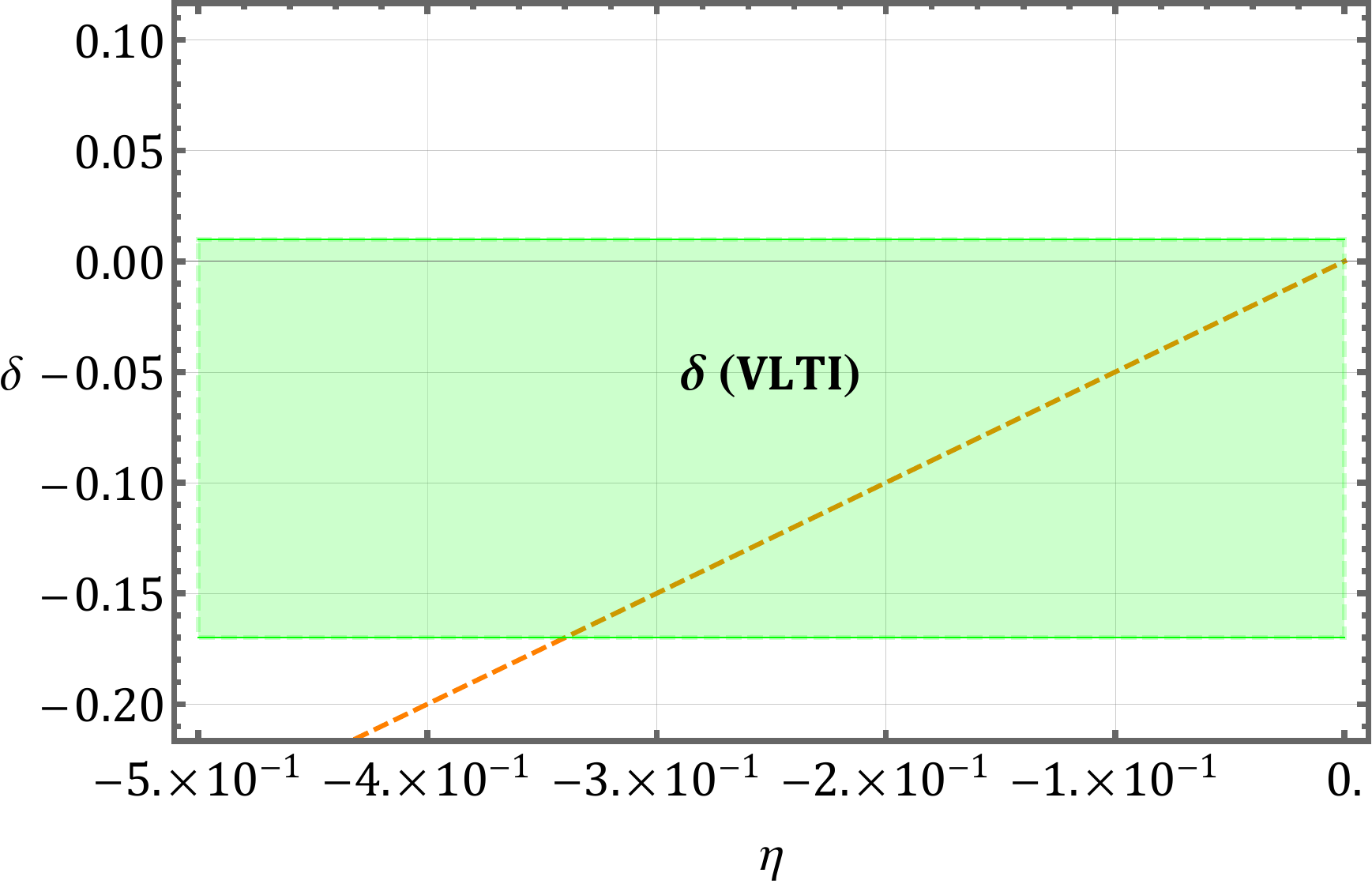}~~~~
		\includegraphics[scale=0.5]{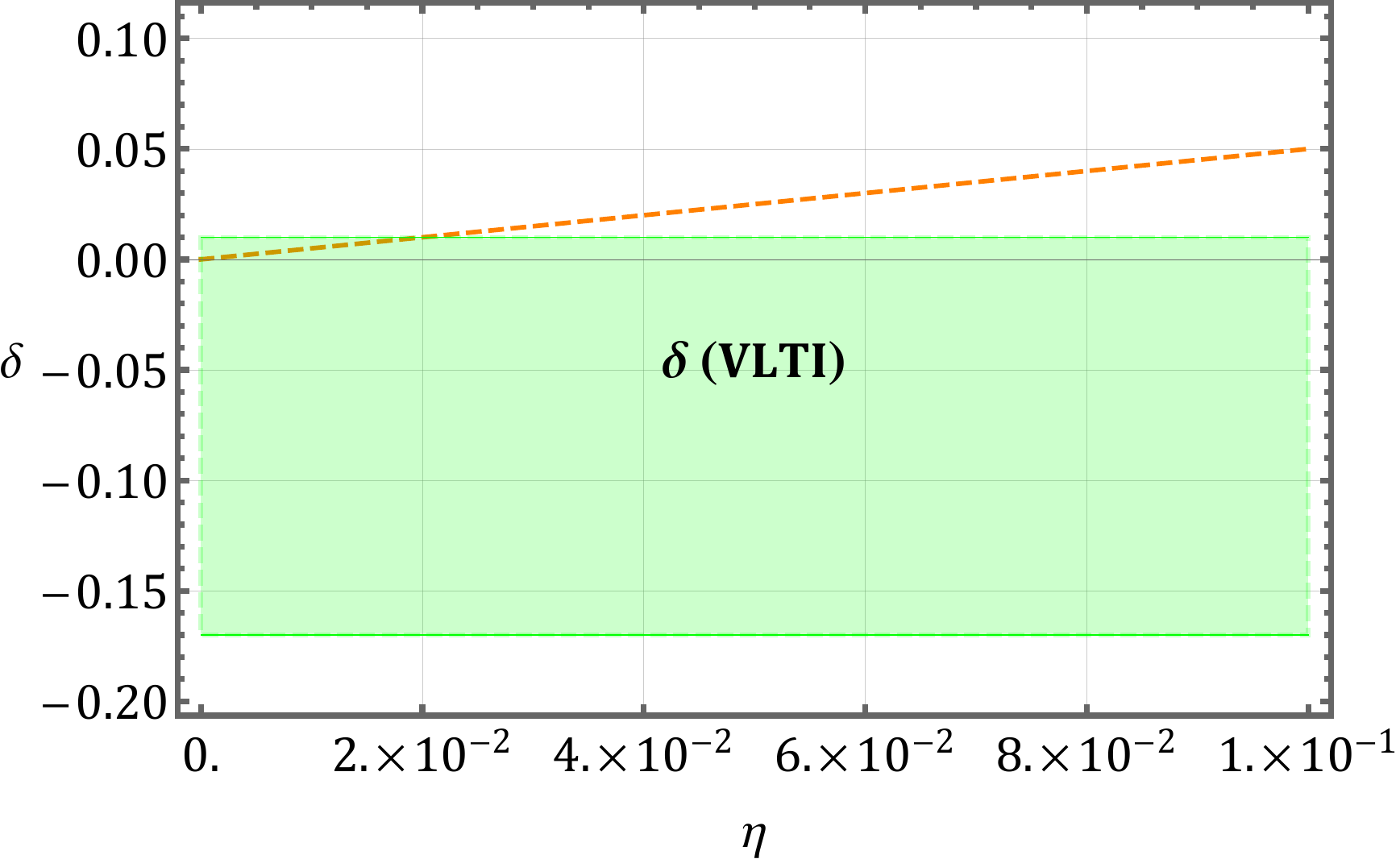}
		\caption{Same as Fig \ref{DD1} but for the Schwarzschild metric corrected by lapse function $f_{II}(r)$ with the LVCs $\eta$.}
		\label{DD2}
	\end{figure}
	
	
	\section{The effect of LV on the shadow size of Schwarzschild black hole}\label{secs.Shadow}
	In this section, we investigate the shadow radius of the Schwarzschild solution corrected with LV terms inspired by SME parameters. Being located on the plane of a faraway observer, the boundary of the black hole shadow marks the apparent image of the photon region  via separating capture orbits from scattering orbits. Concerning the photon region, it is, in essence, the boundary of the region of spacetime that if spherically symmetric, addresses the photon sphere \cite{Cunha:2018acu}.

	Let us begin with the Lagrangian $\mathcal{L}(x,\dot{x})=\frac{g_{\mu\nu}}{2}\dot{x}^{\mu}\dot{x}^{\nu}$
	for the geodesics of spherically symmetric and static spacetime metric
	\begin{equation}
		\begin{aligned}\mathcal{L}(x,\dot{x})=\frac{1}{2}\left(f(r)\dot{t}^{2}-f(r)^{-1} \dot{r}^{2}-r^{2}\left(\dot{\theta}^{2}+\sin^{2}\theta\dot{\phi}^{2}\right)\right)~.\end{aligned}
	\end{equation}
	As usual, by using the Euler-Lagrange equation $\frac{d}{d\lambda}\left(\frac{\partial\mathcal{L}}{\partial\dot{x}^{\mu}}\right)-\frac{\partial\mathcal{L}}{\partial x^{\mu}}=0$
	in the equatorial plane ($\theta=\pi/2)$, the two conserved quantities are the energy $E$ and the angular momentum $L$ \footnote{In case of explicit dependency of the lapse function to $\phi$, the angular momentum is no longer a constant of motion.}, that is
	\begin{equation} \label{cons}
		E=f(r)\dot{t},\quad L=r^{2}\dot{\phi}.
	\end{equation}
	By taking the null-geodesics equation for light, we have 
	\begin{equation}\label{frG}
		f(r)\dot{t}^{2}-f(r)^{-1}\dot{r}^{2}+r^{2}\dot{\phi}^{2}=0\,.
	\end{equation} 
	Inserting the conserved quantities $E$ and $L$, see (\ref{cons}),  into the above equation (\ref{frG}), we arrive at the following orbit equation for photon
	\begin{equation}
		\left(\frac{dr}{d\phi}\right)^{2}=r^{2}f(r)\left(\frac{r^{2}}{f(r)}\frac{E^{2}}{L^{2}}-1\right)~.\label{eff}
	\end{equation}
	Now, we can re-express it in the form of the effective potential 
	\begin{equation}
		\left(\frac{dr}{d\phi}\right)^{2}=V_{eff}~,
	\end{equation}
	with 
	\begin{equation}
		V_{eff}=r^{4}\left(\frac{E^{2}}{L^{2}}-\frac{f(r)}{r^{2}}\right)~.
	\end{equation}
	Given that the orbit equation depends only on the impact parameter $b=L/E$ at the turning point of the trajectory $r=r_{ph}$, we have to demand the conditions $dr/\left.d\phi\right|_{r_{ph}}=0$ or $V_{eff}=0,\quad V_{eff}^{\prime}=0$
	\cite{Ch:book}. This results in the following relation for the impact parameter at the turning point
	\begin{equation}
		b^{-2}=\frac{f(r_{ph})}{r_{ph}^{2}}\,.
		\label{impact}
	\end{equation}
	To obtain the radius of the photon
	sphere $r_{ph}$, one has to impose the conditions $dr/\left.d\phi\right|_{r_{ph}}=0$ and
	$d^{2}r/d\phi^{2}|_{r_{ph}}=0$, leading to the following equations 
	\be\label{photon0}
	\frac{d}{dr}(\frac{r^2}{f(r)})_{r_{ph}}=0~,\\
	\frac{f^{\prime}\left(r_{ph}\right)}{f\left(r_{ph}\right)}-\frac{2}{r_{ph}}=0.\label{photon}
	\ee
	By putting (\ref{impact}) into (\ref{photon0}) and (\ref{photon}), one can easily find
	the location of the photon sphere $r_{ph}$ and the critical
	impact factor $b_{crit}$. As a cross-check, one can show that for the standard Schwarzschild metric, these two are $3M$ and $3\sqrt{3}M$, respectively. Actually, in the spherically symmetric spacetime, the black hole shadow is obtained
	by using the light rings, which correspond to a critical
	points of Eq. (\ref{photon}). As a result, the form of Eq. (\ref{eff}) can be re-expressed as  
	\begin{equation}
		\left(\frac{dr}{d\phi}\right)^{2}=\left(\frac{r^4 f(r_{ph})}{r_{ph}^2 }-r^2f(r)\right).
	\end{equation}
	To calculate the shadow radius $R_{sh}$ from the view of an observer located in $r_0$, it is common to use the angle $\alpha_{\mathrm{sh}}$ between the light ray and the radial direction as follows \cite{Perlick:2021aok}
	\begin{equation}
		\cot\theta_{\mathrm{sh}}=\left.\frac{1}{\sqrt{f(r)r^2}}\frac{dr}{d\phi}\right|_{r=r_{0}}.
	\end{equation}
	and 
	\begin{equation}
		\cot^{2}\theta_{\mathrm{sh}}=\frac{r_0^2 f(r_{ph})}{r_{ph}^2f(r_0)}-1~,
	\end{equation}
	where, by using the relevant trigonometric identities \footnote{$\sin^{2}\theta_{\mathrm{sh}}=(1+\cot^{2}\theta_{\mathrm{sh}})^{-1}$.} and also taking into account $b_{cr}$ of (\ref{impact}), we have 
	\begin{equation}
		\sin^{2}\theta_{\mathrm{sh}}=\frac{b_{\mathrm{cr}}^{2}f\left(r_{0}\right)}{r_0^2}.
	\end{equation}
	The shadow radius of the black hole for a static observer $r_{0}$ is 
	\begin{equation}
		R_{\mathrm{sh}}=r_{0}\sin\theta_{\mathrm{sh}}=\sqrt{\frac{r_{ph}^2f\left(r_{0}\right)}{f\left(r_{ph}\right)}}~,
	\end{equation}
	where for a static observer at far away distance reads as 
	\begin{equation}\label{R}
		R_{\mathrm{sh}}=\frac{r_{ph}}{\sqrt{f\left(r_{ph}\right)}}.
	\end{equation} 
	Since in limit $r_0\rightarrow\infty$, then $f(r_0)\rightarrow1$.
	
	Now we are in a suitable position to reveal the role of the underlying dimensionless LVCs on the size of the shadow radius with spherical symmetry. In Fig. \ref{sh} we plot the shadow resulting from the some optional values of $\xi$ and $\eta$, which respectively address the lapse functions (\ref{LVM2}) and (\ref{LVM1}). As is evident, the presence of LV terms with negative and positive signs in the Schwarzschild metric results in the decrease and growth of the shadow size, respectively. Here, $\eta$ has a larger contribution to the change of the shadow size, compared to $\xi$. Note that the values fixed for $\xi$ and $\eta$ in Fig. (\ref{sh}), are not realistic. Indeed, they merely have been used to reveal the impact of LVCs on the size of the black hole shadow, and in what follows, their allowed ranges in light of the data related to Sgr A* will be derived.

	\section{Constraints on the SME Coefficients from Sgr A* compact object}\label{secs.sgr}
	
	In this section we derive, by using data released by EHT for Sgr A*, some allowed ranges for two combinations of LVCs $\xi$ and $\eta$. We perform this analysis by comparing the observed angular radius of the ring-like in the EHT near-horizon image of Sgr A* with the SME inspired Schwarzschild metric at hand. In this way, the condition of compatibility within the range of uncertainties allowed by the EHT, permits us to set constraints on the $\xi$ and $\eta$. To this end, two ingredients are essential. 
	The first is the mass-to-distance ratio of Sgr A*, $M/D$. 
	By exploiting the stellar cluster dynamics, and the motion of S-stars, 
	the mass and distance to Sgr A* have been extracted by two teams ``Keck'' \cite{Do:2019txf}, and ``VLTI'' \cite{GRAVITY:2020gka} as follows
	\begin{align} \label{KV}
		&M=(3.951 \pm 0.047)\times 10^6 M_\odot~,~~~~~~~~~~~~~~~D=(7.953\pm 0.050\pm0.032)~ \mbox{Kpc}~, \nonumber \\
		&M=(4.297 \pm 0.012\pm0.040)\times 10^6 M_\odot~,~~~~D=(8.277\pm 0.009\pm0.033)~ \mbox{Kpc}~,
	\end{align} respectively. 
	The second is a calibration factor, connecting the size of the bright ring of emission to the size of the corresponding shadow. This, in essence, quantifies the level of safety to use the size of the bright ring as a proxy for the shadow size. Avoiding the details, by folding in the calibration factor with uncertainties in the mass-to-distance ratio of Sgr A*, and also taking into account the angular diameter of the bright ring of Sgr A*, one obtains the fractional deviation $\delta$ between the inferred shadow radius of EHT and the Schwarzschild one. The fractional deviation from the Schwarzschild expectation, $\delta$, depends on the mass-to-distance ratio. 
	So, based on the Keck and VLTI measurements reported above, one can record the following values for $\delta$ \cite{EventHorizonTelescope:2022gsd}
	\be\label{de}
	\delta=-0.04^{+0.09}_{-0.10}~~~\mbox{(Keck)},~~~~~\delta=-0.08^{+0.09}_{-0.09} ~~~\mbox{(VLTI)}
	\ee 
	Interestingly, we see $\delta<0$, meaning that there is very slight preference for the shadow of Sgr A* being slightly smaller than the Schwarzschild one. Now by taking into account of the average of the Keck and VLTI-based estimates of $\delta$  \cite{Vagnozzi:2022moj}
		\begin{eqnarray}
			\delta\simeq-0.06 \pm 0.065\,,
			\label{eq:av}
		\end{eqnarray}	
		one obtains the following shadow radius for Sgr A* \cite{Vagnozzi:2022moj}
		\begin{eqnarray}
			4.54 \lesssim \frac{R_{\rm Sgr A*}}{M}\lesssim 5.22\,,
			\label{eq:1sigma}
		\end{eqnarray}
		within $1\sigma$ uncertainty. In what follows, with this idea in mind that the Sgr A* compact object recorded by EHT can be described by the SME inspired Schwarzschild metric
		at hand, and using (\ref{eq:1sigma}) and (\ref{eq:av}), we will extract some constraints for two combinations of the LVCs $\xi$ and $\eta$.
	
	By using the shadow radius (per mass) computed in the previous section (see Eq. (\ref{R})), we plot in Fig. \ref{RR} the radius of the resulting black hole shadow as a function of the LVCs $\xi$ and $\eta$, together with $1\sigma$ uncertainty on the radius of the shadow of Sgr A* as reported in (\ref{eq:1sigma}).
		Equivalently, in the light of the average obtained from Keck and VLTI measurements of fractional deviation i.e., (\ref{eq:av}), we perform this analysis in Fig. \ref{DDD} for the fractional deviation predicated by the SME-inspired Schwarzschild with the laps functions  (\ref{LVM2}), and (\ref{LVM1}), respectively.
		As expected, we see a decrease and growth in the shadow radius of Schwarzschild in the presence of negative and positive LVCs, respectively. This trend corresponds to $\delta<0$ and $\delta >0$ in 	 Fig. \ref{DDD}, respectively.  Concerning the positive LVCs, we clearly see from  Fig. \ref{RR} (also \ref{DDD}) that the shadow radius derived for Sgr A* by EHT restricts the growth of the shadow size of the underlying black hole such that values beyond $\xi \sim 1.5 \times 10^{-2}$ and $\eta \sim 10^{-2}$ are ruled out. This results in upper bounds $ \lesssim 1.5 \times 10^{-2}$ and $\lesssim 10^{-2}$, for $\xi$ and $\eta$, respectively. For negative LVCs, these two figures reveal lower bounds $\xi>-3.5\times 10^{-1}$ and $\eta\gtrsim -2.5\times 10^{-1}$ since in this case, the shadow radius of Sgr A* restricts the reduction of the shadow size. As a supplementary analysis, in Figs. \ref{DD1} and \ref{DD2} we repeat Fig. \ref{DDD} for each of  Keck and VLTI measurements of fractional deviation, separately. It is clear from these figures that a more robust constraint than the previous ones will not be obtained.
		%

	An important point that we have to note about the aforementioned upper bounds is that one cannot realize their origin.
	Namely, it may not be clear whether the LV correction results in changes in the shadow size or whether it is related to the masses measured for Sgr A* from two different sources:  Keck and VLTI. In other words, one can simultaneously attribute the effect of LV to the shadow radius and the measured mass since uncertainty in both is of the same order of magnitude. The reason behind it is that the fractional deviation from the Schwarzschild metric, which, in essence, comes from the shadow radius, has a high dependency on the mass-to-distance ratio. Such a degeneracy situation is not threatening the validation of the upper bounds since, in any case, they are obtained from one of the properties belonging to Sgr A*'s image, whatever the size of the shadow radius or the mass.

	\section{discussion and Conclusion}\label{secs.con}
	The extraordinary images of supermassive objects in the center of galaxies, M87* and Milk Way (Sgr A*) which were released by the Event Horizon Telescope (EHT),  respectively on 10 April 2019 and 12 May 2022, give us a novel possibility of shedding light on metric theories of gravity as well as the fundamental physics in the strong-field regime. Despite some theoretical challenges, these two compact objects seem to be strong candidates for playing the role of the black hole in nature. Besides, the resolution of EHT measurements is limited, and despite a good consistency reported between the Sgr A* 's shadow and prediction of the general theory of relativity (GTR), it can be seen as a tiny space that potentially can be controlled by adding some fundamental corrections into the metric.
	One of the fundamental issues in physics which plays a key role in the standard model of particle physics as well as GTR is the Lorentz symmetry. Different approaches leading to quantum gravity, and unified theories are the key motivations behind significant efforts to test Lorentz symmetry, and eventually its possible breakdown at some scales.

	In this paper, we have used the near-horizon images of Sgr A*, recently captured by EHT to test the Lorentz symmetry violation (LSV) in gravity. A comprehensive framework to address it is 
	an effective field theory well known as the Standard-Model Extension (SME) in which are included all possible Lorentz violating (LV) operators to the standard model and GTR, as well. By adopting the pure gravity sector of the minimal version of the SME, we, in essence, deal with a matrix $4 \times 4$ of the coefficients $\bar{s}^{\mu\nu}$ that describe dominant observable deviations from Lorentz invariance. 
	
	As is common in metric theories of gravity, to infer a constraint on the LSV in gravity by the information released in the first image of Sgr A*, it is essential to have a metric that includes the LV terms. 
	However, within the general SME framework, except for the  bumblebee gravity as a subclass model of SME, there is no exact black hole solution. In the framework of the pure-gravity sector of the minimal SME theory, and using the effective Newtonian potential obtained from the parameterized post-Newtonian (PPN) approximation for a point particle moving slowly, we have derived a Schwarzschild-like metric in a weak gravitational field. It is well-known that any extended theory of gravitation can yield corrections to Newton's potential, such that the PPN formalism could furnish tests for the relevant theory.
	The correction terms in the underlying metric appear as two different combinations of spatial diagonal Lorentz violating coefficients (LVCs) $\xi=\bar{s}^{ii}\equiv\bar{s}^{XX}+\bar{s}^{YY}+\bar{s}^{ZZ}$ and $\eta=\xi-\bar{s}^{ZZ}\equiv\bar{s}^{XX}+\bar{s}^{YY}$.	Recall two useful points. First, due to the spherical symmetry and computing the relevant shadow in the equatorial plane ($\theta=\pi/2$), the non-diagonal components of LVCs in the metric, are out of reach. Second, despite the absorption of the temporal diagonal component of LVC ($\bar{s}^{TT}$) into the gravitational constant, the traceless condition $s^{\mu}_{{\phantom \mu} \mu}=0$ (i.e., $\bar{s}^{TT}-\xi=0$), guarantees that the upper bounds, obtained to the spatial diagonal combination $\xi$, can be safely attributed to the time-time coefficient $\bar{s}^{TT}$, too.

	Observations under our attention of the first image of Sgr A*, that have been considered in our analysis, are: the shadow radius per mass $\frac{R_{\rm Sgr A*}}{M}$, and the fractional deviation from standard Schwarzschild $\delta$. The former comes from bright ring diameter detected around Sgr A*, and the latter originates from comparing the shadow diameter to the stellar-dynamical measurements of the mass of Sgr A*, i.e., mass measured with the Keck and VLTI  instruments via star orbits (particularly the motion of S-stars). In the light of both observations we have extracted lower and upper bounds: $\xi\gtrsim -3.5\times 10^{-1},~~\eta \gtrsim -2.5 \times 10^{-1}$ and $\xi\lesssim 1.5\times 10^{-2},~~\eta \lesssim 10^{-2}$, respectively.
		The best upper bound for LVCs at hand obtained from two Sgr A* 's observations mentioned above is, hence, at the $10^{-2}$ level. This means that the first image of Sgr A* recorded by EHT does not permit us to probe LSV with a sensitivity level more than one per hundred. 
	However, the form of the correction of LSV in the metric causes a dilemma, meaning that the origin of these upper bounds comes from the measurement of Sgr A*'s mass by different sources or is related to the shadow size of Sgr A*. It is for that the uncertainty measurements of mass and shadow radius of Sgr A* is of the same order of magnitude. More precisely, the fractional deviation from the Schwarzschild metric, in essence, comes from the shadow radius and has a high dependency on the mass-to-distance ratio.	Overall, it does not matter since, in any case, they are obtained from one of the properties belonging to Sgr A*'s image, whether the size of the shadow radius or mass.

	By comparing with constraints derived from some of the well-known frameworks (listed in the Introduction), one finds that relative to most cases, the sensitivity level of the existing upper limit is quantitatively weaker. 
	The best constrain from Sgr A* with the present resolution (one per hundred) is weaker than most other sensitivities in search for LSV. It is necessary to stress that one should interpret this upper bound differently from other published ones since, in the standard SME studies, the Sun-centered celestial frame is chosen as the standard frame, while this is not the case here. In other words, although the best upper bound is weaker than all limits listed in the Introduction, it is obtained from the scan of gravitational SME in the strong-field regime such as around the black hole horizon.
	Apart from this, the need to increase the types of measurements is inevitable, even if those do not give us more strong constraints. In other words, every novel window opened to us in the universe is potentially prone to shedding light on the validity of fundamental physics. Even though the best upper limit obtained from the first image of Sgr A* is not so strong to compete with other setups, it is valuable for two reasons. First, it is an independent constraint arising from data of the near-horizon scales, which is a gravity-dominated region. 	
	Second, despite the existing metric, in essence, does not allow us to probe the SME in a strong-field regime around the horizon, these upper bounds are merely a promising first step. In other words, one can hope for more sensitive measurements of the deviation of the inferred shadow image of Sgr A* from Schwarzschild in the future, resulting in more stringent upper limits on SME coefficients. Namely, if the mentioned deviation with much greater sensitivity is measured, even with this weak-field metric, one can extract tighter constraints on the underlying LVCs. 
	
	\vspace{0.5cm}
	{\bf Acknowledgments:}
	We sincerely appreciate Alan Kostelecky and Quentin Bailey for carefully reading and constructive comments on the manuscript. We also thank Marco Schreck for helpful discussions and points.
	M. Kh thanks Carlos Herdeiro and Sunny Vagnozzi for enlightened discussions on some information published of the Sgr A*'s image. We thank the anonymous referees for the valuable comments that helped us improve this manuscript.
	
	%
	
\end{document}